\documentclass[aps,prx,twocolumn,superscriptaddress]{revtex4-2}

\usepackage[utf8]{inputenc}
\usepackage{graphicx}%
\usepackage{siunitx}
\usepackage{multirow}%
\usepackage{amsmath,amssymb,amsfonts}%
\usepackage{amsthm}%
\usepackage{mathrsfs}%

\begin{document}

\title{Two-Photon Interference of Photons from Remote Tin-Vacancy Centers in Diamond}

\author{Vladislav Bushmakin}
\email{v.bushmakin@pi3.uni-stuttgart.de}
\affiliation{3.\ Physikalisches Institut, Universit\"at Stuttgart, Stuttgart 70569, Germany}
\affiliation{Max Planck Institute for Solid State Research, Stuttgart 70569, Germany}

\author{Oliver von Berg}
\affiliation{3.\ Physikalisches Institut, Universit\"at Stuttgart, Stuttgart 70569, Germany}

\author{Colin Sauerzapf}
\affiliation{3.\ Physikalisches Institut, Universit\"at Stuttgart, Stuttgart 70569, Germany}

\author{Sreehari Jayaram}
\affiliation{3.\ Physikalisches Institut, Universit\"at Stuttgart, Stuttgart 70569, Germany}

\author{Andrej Denisenko}
\affiliation{3.\ Physikalisches Institut, Universit\"at Stuttgart, Stuttgart 70569, Germany}

\author{Cristina Tarín}
\affiliation{Institute for System Dynamics, Universit\"at Stuttgart, Stuttgart 70563, Germany}

\author{Jens Anders}
\affiliation{Institute of Smart Sensors, Universit\"at Stuttgart, Stuttgart 70569, Germany}

\author{Vadim Vorobyov}
\affiliation{3.\ Physikalisches Institut, Universit\"at Stuttgart, Stuttgart 70569, Germany}

\author{Ilja Gerhardt}
\affiliation{Institut für Festkörperphysik, Leibniz Universität Hannover, Hannover 30167, Germany}

\author{Di Liu}
\affiliation{3.\ Physikalisches Institut, Universit\"at Stuttgart, Stuttgart 70569, Germany}
\affiliation{John A. Paulson School of Engineering and Applied Sciences, Harvard University, Cambridge, MA 02138, USA}

\author{Jörg Wrachtrup}
\affiliation{3.\ Physikalisches Institut, Universit\"at Stuttgart, Stuttgart 70569, Germany}
\affiliation{Max Planck Institute for Solid State Research, Stuttgart 70569, Germany}

\begin{abstract}
Scalable quantum networks rely on optical connections between long-lived qubits to distribute entanglement.
Tin vacancies in diamond have emerged as promising qubits, 
offering extended spin coherence times at liquid helium temperatures and spin-dependent, 
highly coherent optical transitions for effective photon-based communication.
Connecting remote nodes requires quantum interference of indistinguishable photons, 
which is challenging in an inhomogeneous solid-state environment.
Here, we demonstrate a two-node experiment with tin vacancies in diamond, 
which exhibit a resonant frequency distribution spanning approximately 8~GHz.
To overcome the frequency mismatch, we tune the resonant frequencies of 
one node using the Stark effect. We achieve tunability up to \SI{4}{\giga\hertz} 
while maintaining optical coherence.
As a demonstration, we achieve detuning-dependent remote two-photon interference 
between separate nodes, obtaining \SI{80+-6}{\percent} interference visibility 
without postprocessing when the defects' optical transitions are tuned into resonance, 
and \SI{63+-8}{\percent} with detuning up to 20 times their natural linewidths.
These results highlight the potential of tin-vacancy centres in diamond for establishing 
robust optical links between remote quantum registers.
\end{abstract}

\maketitle

\section{Introduction}\label{sec:introduction}

Developing scalable quantum networks, where local quantum processing nodes are interconnected via indistinguishable photons, remains a central challenge in quantum information science~\cite{awschalom_development_2021}. 
Reliable optical links must be efficiently interfaced 
  with quantum memories for such networks to function. 
  Significant progress has been achieved with various platforms, including 
  trapped atoms and ions~\cite{ritter_elementary_2012, hofmann_heralded_2012, bock_high-fidelity_2018, reiserer_quantum_2014, hucul_modular_2015, rosenfeld_event-ready_2017}, 
  quantum dots~\cite{zaporski_ideal_2023, yu_telecom-band_2023}, organic molecules~\cite{siyushev_molecular_2014},
  rare-earth-doped solids~\cite{lago-rivera_telecom-heralded_2021,ruskuc2024scalable}, transmon qubits~\cite{kurpiers_deterministic_2018}, and solid-state defects~\cite{hensen_loophole-free_2015, awschalom2018quantum,higginbottom2022optical, morioka2020spin, fang2024experimental}.
  Among these systems, color centers in diamond have emerged as promising candidates for scalable quantum applications due to their compatibility 
  with nanophotonic devices~\cite{nguyen2019integrated, knaut_entanglement_2024} and their long spin coherence times~\cite{sukachev2017silicon, stas_robust_2022}.
  \begin{figure}[ht] 
  \centering 
  \includegraphics[width=0.5\textwidth]{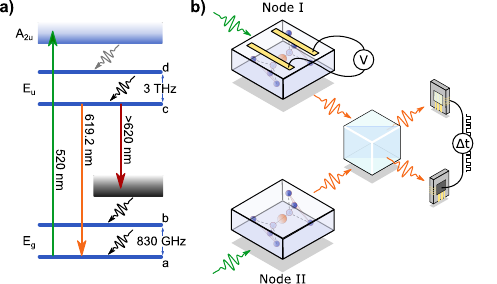} 
  \caption{(a) Energy level diagram for the negatively charged tin-vacancy (SnV\textsuperscript{--}) 
  centre in diamond, illustrating the electronic ground and excited states with their respective orbital splittings. 
   The c-a transition (\SI{619.2}{\nano\meter}) between the lowest orbital levels is used in this experiment. 
   (b) Concept of the remote two-photon experiment with one of the nodes tuned with an electric field.
} 
  \label{fig:Fig_I} 
\end{figure}
  Group-IV vacancy centers in diamond, such as the silicon-vacancy (SiV\textsuperscript{--}) centre, 
  offer distinct advantages owing to their inversion symmetry~\cite{hepp_electronic_2014,thiering_ab_2018}, which renders them less 
  sensitive to electric field fluctuations. This property enables integration into nanostructures 
  without degrading optical coherence~\cite{sipahigil_integrated_2016}. 
  SiV\textsuperscript{--} centers have demonstrated narrow inhomogeneous distributions, 
  Fourier-limited linewidths~\cite{evans2016narrow, schroder2017scalable}, and stable optical transitions, facilitating two-photon interference 
  experiments between remote emitters~\cite{sipahigil_indistinguishable_2014, bhaskar_experimental_2020}. 
  However, SiV\textsuperscript{--} centers suffer from low quantum efficiency and require operation at sub-Kelvin 
  temperatures to achieve long spin coherence times~\cite{jahnke_electronphonon_2015, sukachev_silicon-vacancy_2017}, 
  which is impractical for scalable quantum networks.

  Heavier group-IV defects, such as germanium-vacancy (GeV\textsuperscript{--}), 
  tin-vacancy (SnV\textsuperscript{--}), and lead-vacancy (PbV\textsuperscript{--}) centers, 
  offer enhanced properties, including higher quantum efficiency, larger spin-orbit splitting, 
  and longer spin relaxation times at elevated temperatures 
 ~\cite{siyushev_optical_2017, iwasaki_tin-vacancy_2017, trusheim_transform-limited_2020, senkalla_germanium_2024,zahedian_blueprint_2024}. 
  Notably, the SnV\textsuperscript{--} (see energy level diagram in Figure~\ref{fig:Fig_I}a) centre has demonstrated spin-lattice relaxation times $T_1$ reaching 1~s at temperatures above 1~K and coherence times $T_2$ 
  extending up to 10~ms with dynamical decoupling 
 ~\cite{guo_microwave-based_2023, rosenthal_microwave_2023, karapatzakis_microwave_2024}. 
  Moreover, scalable integration of these centers onto CMOS-compatible photonic chips 
  further positions the SnV\textsuperscript{--} centre as a promising candidate for quantum 
  networking applications at practical temperatures~\cite{li_heterogeneous_2024}.
 
  Despite these promising characteristics, it remains challenging to achieve high optical coherence 
  to perform two-photon interference with SnV\textsuperscript{--} centers. 
  In solid-state environments like diamond, variations in the local environment surrounding 
  each defect lead to inhomogeneous broadening — a spread in the distribution of the defects' 
  zero-phonon line (ZPL) frequencies. For SnV\textsuperscript{--} centers, the 
  larger atomic size of the tin ion exacerbates spectral broadening, as implantation-induced vacancies can 
  introduce significant lattice strain and 
  electric fields, resulting in broad ZPL distributions 
  ranging from tens to hundreds of gigahertz~\cite{trusheim_transform-limited_2020}. 
  Recent studies have demonstrated that it is possible to reduce the 
  inhomogeneous broadening of SnV\textsuperscript{--} centers to approximately 
  4~GHz by employing HPHT annealing~\cite{narita_multiple_2023}. Although 
  this improvement marks significant progress, the reduced inhomogeneous 
  broadening still remains over 130 times larger than the natural linewidth, 
  making it extremely unlikely to find two defects with overlapping emission 
  frequencies for interference experiments. Consequently, tuning mechanisms 
  are required to bring individual SnV\textsuperscript{--} centers into resonance. 
  Previous efforts to tune optical transitions through application of electric fields have been limited, as they cause significant spectral 
  diffusion~\cite{aghaeimeibodi_electrical_2021, de_santis_investigation_2021}. 
  Spectral diffusion degrades the optical coherence necessary for 
  high-visibility two-photon interference, presenting a significant 
  obstacle to utilizing SnV\textsuperscript{--} centers in scalable quantum networks.

  In this work, we address the challenges of achieving high optical coherence and performing 
  two-photon interference with SnV\textsuperscript{--} centers with the diamond chips in two remote optical setups separated by 70~meters of optical fiber (Node I and Node II, depicted conceptually in Figure~\ref{fig:Fig_I}b and in detail in Figure~\ref{fig:Fig_III}a). 
  We create SnV\textsuperscript{--} centers in diamond with low strain and 
  a narrow inhomogeneous distribution. By implementing Stark tuning of 
  optical transitions with minimal decoherence, we bring the optical transitions 
  of separate defects into resonance. 
  By exciting with off-resonant green 520 nm laser and filtering the narrow c-a transition (see Figure~\ref{fig:Fig_I}a) we perform the two-photon 
  interference of single photons from remote  SnV\textsuperscript{--} and probe the optical indistnguiability at various detunings.

\section{Results}\label{sec:results}
\subsection*{Spectral distribution of the resonant transitions in low-strain samples}
\begin{figure*}[ht]
  \centering
  \includegraphics[width=1\textwidth]{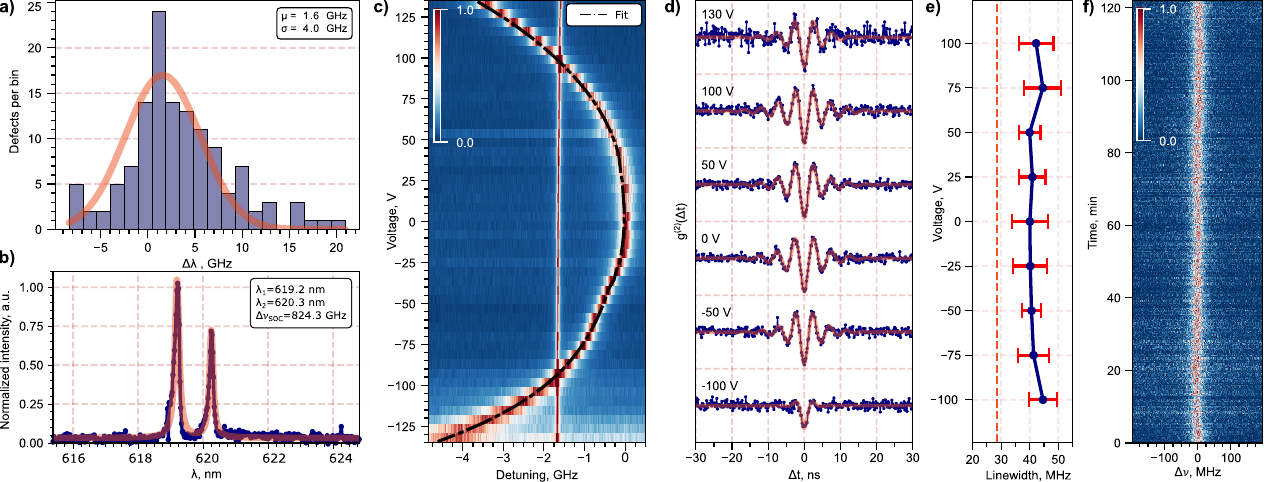}
  \caption{Statistics of optical properties of obtained colour centers. (a) Histogram of the ZPL distribution for SnV\textsuperscript{--} centers implanted at \SI{40}{\kilo\electronvolt}, 
  with a fitted Gaussian curve showing an inhomogeneous broadening of $\sigma = \SI{4.0}{\giga\hertz}$. 
  The centre of the distribution is offset by $\mu = \SI{1.6}{\giga\hertz}$ from \SI{484.130}{\tera\hertz}.
   (b) Photoluminescence (PL) spectrum showing the distribution of zero-phonon lines (ZPLs) for 110 defects, with two prominent 
   peaks at \SI{619.2}{\nano\meter} and \SI{620.3}{\nano\meter}, corresponding to \SI{824.3}{\giga\hertz} orbital splitting and reduced 
   inhomogeneous broadening, indicating the low-strain environment in ion-implanted samples.
  (c) Detuning of the optical transition as a function of voltage, showing a tuning range of up to \SI{4}{\giga\hertz}. 
  The straight line is the signal from the second defect in the remote node, showing the control over the spectral overlap.
  The black dashed curve is the fit function obtained by introducing a two-level system (a charge trap) interacting with the SnV\textsuperscript{--}.
  (d) The autocorrelation function $g^{(2)}(\Delta t)$ displays Rabi oscillations at different voltages, 
  indicating that coherent oscillations persist despite Stark tuning.
  (e) Tuning of the optical transition of the tin-vacancy (SnV\textsuperscript{--}) 
  centre using an applied electric field at voltages from $\SI{-100}{\volt}$ to $\SI{100}{\volt}$. 
  The linewidths remain near Fourier-limited up to a \SI{1}{\giga\hertz} tuning range. 
  The red dashed line on the side plot marks the Fourier limit of (\SI{28.4}{\mega\hertz}) for an emitter with $T_1 = \SI{5.6}{\nano\second}$ lifetime. 
  (f) The optical transition demonstrated spectral stability, showing a single-scan linewidth of \SI{31.5}{\mega\hertz} 
  and an averaged linewidth of \SI{31.9}{\mega\hertz} over a 2-hour measurement period.}
  \label{fig:Fig_II}
\end{figure*}

To minimize lattice damage during defect creation, we implanted ions into the samples along crystal channels, 
aligning the ions at a 0$^\circ$ incidence angle relative to the [001] 
diamond surface~\cite{lehtinen_molecular_2016, santonocito_suppression_2024}. 
This approach reduces implantation-induced defects and yields 
favorable optical properties across a range of implantation energies, with 
lower energies resulting in the least lattice damage. Figures~\ref{fig:Fig_II}a and \ref{fig:Fig_II}b 
show results from the \SI{40}{\kilo\electronvolt} implanted sample, 
where we observe a ground state splitting value of \SI{824}{\giga\hertz} and a narrow spectral distribution of zero-phonon 
lines (ZPLs) with a linewidth variation of $\sigma\approx\SI{4.0}{\giga\hertz}$, 
due to a low-strain environment. This sample thus serves as a best-case scenario, 
exhibiting minimal spectral broadening. For the remaining experiments presented in this study, 
we focus on samples implanted at \SI{80}{\kilo\electronvolt} and \SI{170}{\kilo\electronvolt}, 
as the increased implantation depth places the defects farther from the diamond surface, 
reducing photoinduced 
spectral drift effects caused by the strong off-resonant excitation. 
A forthcoming study will present a 
comprehensive statistical analysis of channeled implantation across different energies.

\subsection*{Coherent properties of the electronic transitions under Stark tuning}

\begin{figure*}[ht]
  \centering
  \includegraphics[width=1\textwidth]{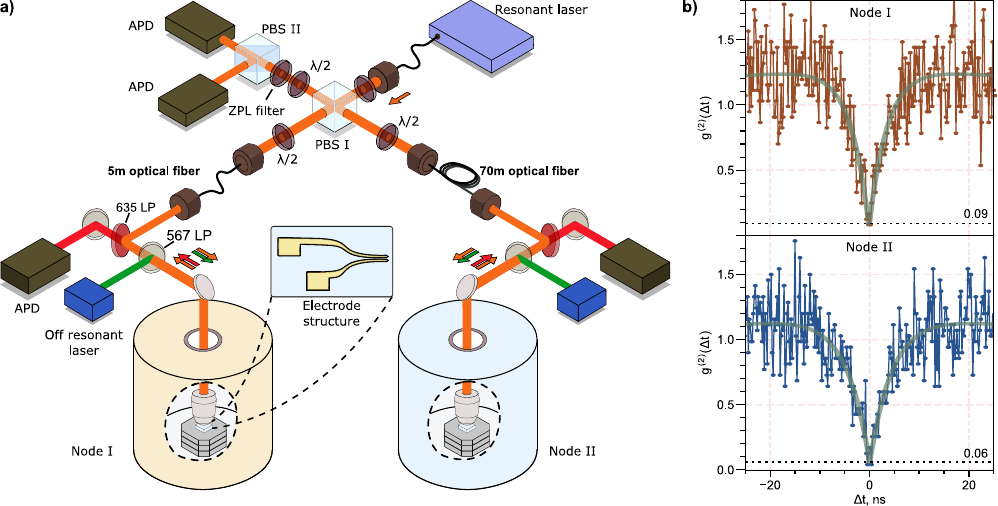}
  \caption{Two-node tin-vacancy experiment setup. (a) Schematic of the optical setup.
  The system consists of two cryostats, node I
  and node II, located in separate rooms and connected by a \SI{70}{\meter} 
  long polarization-maintaining (PM) fiber. 
  Off-resonant excitation using a \SI{520}{\nano\meter} laser was employed, 
  with the emission from the tin-vacancy centers filtered by 
  long-pass filters and guided to an interferometer via the fibers. Half-wave 
  plates and polarizing beam splitters were used to control and 
  combine the emission from both nodes. The combined output is filtered with an $\approx$~\SI{0.5}{\nano\meter} broad
  bandpass filter, and the 
  resulting signal is detected using avalanche photodiodes to measure coincidence counts. 
  (b) Autocorrelation 
  function $g^{(2)}(\Delta t)$ measured for single defects in both cryostats, confirming that each 
  node contains a single defect 
  with $g^{(2)}_{\mathrm{I}}(0) \approx 0.09$ and $g^{(2)}_{\mathrm{II}}(0) \approx 0.05$.} 

  \label{fig:Fig_III}
\end{figure*}

To enable tuning of defects to account for inhomogeneous broadening, we 
fabricated a pair of parallel electrodes to create a controlled electric field (see Methods and Figure~\ref{fig:Fig_III}a), 
allowing us to tune the optical transitions of 
SnV\textsuperscript{--} centers.

In contrast to prior work by De Santis et al.~\cite{de_santis_investigation_2021}, 
we identified SnV\textsuperscript{--} centers that maintain optical coherence across a 
tuning range of up to \SI{4}{\giga\hertz}, as demonstrated by optical Rabi 
oscillations. Figure~\ref{fig:Fig_II}d shows the autocorrelation functions $g^{(2)}(\Delta t)$ 
measured across applied voltages from \SI{-100}{\volt} to \SI{130}{\volt} 
under resonant excitation with a continuous-wave charge repump laser. 
Fitting these measurements with a Rabi oscillation model~\cite{wang_optical_2022} 
shows preserved optical coherence across a tuning range of \SI{4}{\giga\hertz}, 
as demonstrated in Figure~\ref{fig:Fig_II}c.

The kink in the Stark shift curve at approximately \SI{-50}{\volt} (Figure~\ref{fig:Fig_II}c)
 can be attributed to nearby single or multiple charge traps whose occupancy changes under 
the applied electric field. 
As the field strength modulates the traps' energy levels, 
it reaches a threshold where the traps either capture or release charge carriers, 
altering the local electric field at the SnV\textsuperscript{--} centre and causing 
the observed discontinuity. This switching behavior is further influenced by the traps' population 
fluctuations at higher fields, leading to additional spectral broadening 
\cite{segura_tip-induced_2001, bassett_electrical_2011}.
We model the charge-trap effect with two fourth-order polynomials, 
corresponding to the traps being filled or empty, as shown by the black dashed fit in Figure~\ref{fig:Fig_II}c.

To further examine the nature of the spectral broadening, we conducted resonant excitation experiments 
at low powers, omitting the use of the charge repump laser, which can facilitate 
photoionization of charge traps and induce spectral diffusion~\cite{bassett_electrical_2011, gorlitz_spectroscopic_2020}. 
By using only resonant excitation, we preserved near-Fourier-limited linewidths of 
approximately \SI{40}{\mega\hertz}, within $\approx$1.5 times the Fourier limit. 
The optical linewidths can be further narrowed by decreasing 
the temperature~\cite{wang_transform-limited_2024}. We confirmed this by performing 
repeated scans with enhanced cooling and thermal shielding in another cryostat, 
yielding narrower lines from SnV\textsuperscript{--} centers located between the electrodes (Figure~\ref{fig:Fig_II}f).
The optical transition remains spectrally stable with a single-scan 
linewidth of \SI{31.5}{\mega\hertz}
and an averaged linewidth of \SI{31.9}{\mega\hertz} over the 2-hour measurement, 
despite the charge state being reinitialized with the continuous-wave (cw) \SI{520}{\nano\meter} laser.

\subsection*{Detuning-dependent remote two-photon interference}

\begin{figure*}[ht]
  \centering
  \includegraphics[width=1\textwidth]{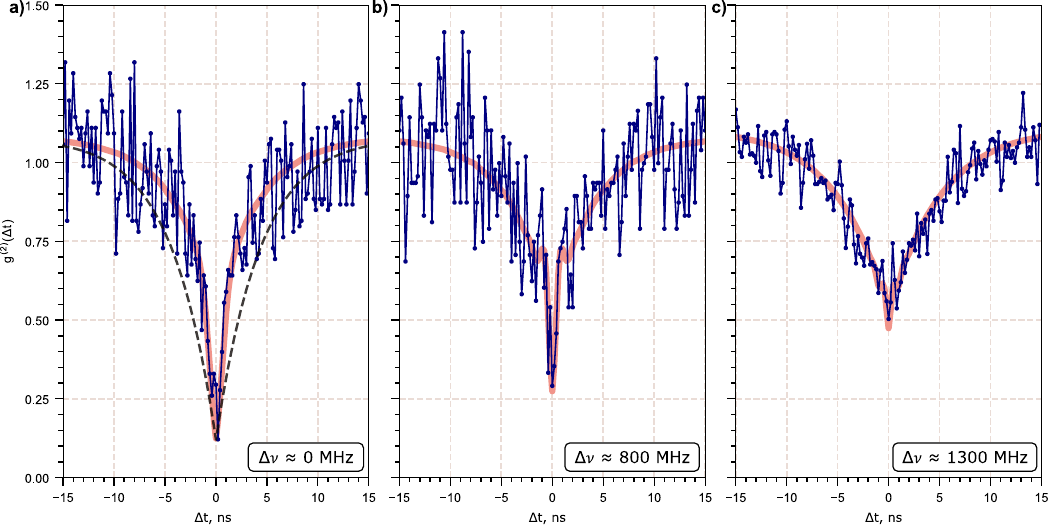}
  \caption{Detuning-dependent two-photon interference of single photons emitted from two remote tin-vacancy centers.
  The second-order cross-correlation function $g^{(2)}(\Delta t)$ is measured at three detuning points: 
  (a) resonant condition with $\Delta \nu \approx 0$~MHz, (b) moderate detuning with $\Delta \nu \approx 800$~MHz, and 
  (c) maximum detuning with $\Delta \nu \approx 1300$~MHz. 
  The blue points represent the recorded data. The red curve is the fit with equation~\ref{eq:g2_sd} for two-photon interference for indistinguishable sources with varying detunings and spectral diffusion.
  The black dashed curve in (a) represents the simulation of the ideal resonant case using the same parameters as the fit to the experimental data, except with zero spectral diffusion.
  The observed Hong-Ou-Mandel dips reveal the two-photon interference even at large detunings, 
  with narrowing interference signatures due to spectral diffusion. Despite the induced spectral diffusion, the interference visibility in resonance 
  was measured to be$ V_{\mathrm{HOM}}(0) = 0.80 \pm 0.06 $ and $ V_{\mathrm{HOM}}(0) = 0.63 \pm 0.08 $ at moderate detuning. 
  At maximum detuning, the interference features are washed out due to the combined effects 
  of large detuning and enhanced spectral diffusion.}
  \label{fig:Fig_IV}
\end{figure*}

The broad tuning range and preserved optical coherence allow us to investigate 
two-photon interference~\cite{hong_measurement_1987, shih_new_1988} between spatially separated SnV\textsuperscript{--} centers (see Figure~\ref{fig:Fig_III}a) 
under frequency detuning via the Stark effect.

To capture the detuning-dependent nature of the two-photon interference, the experiment 
was performed at three voltage setpoints: resonant, moderately detuned, and maximally detuned. 
For the resonant measurements, the setpoint was selected on a slope of the tuning curve 
to enable periodic overlap checks of the resonances. Every five minutes, off-resonant 
excitation was swapped to resonant for overlap verification. The count rate of the zero-phonon 
lines at the interferometer was set by adjusting the excitation powers to yield 
\SI{4}{\kilo cts\per\second} for each node. The measured second-order autocorrelation functions ($g^{(2)}(0)$)
show the purity of the single photon emission (see Figure~\ref{fig:Fig_III}b) from the SnV\textsuperscript{--} used in this experiment.

The \SI{9}{\hour} zero detuning measurement presented in Figure~\ref{fig:Fig_IV}a reveals a 
Hong-Ou-Mandel dip with visible narrowing of the antibunching compared to an ideal case with zero spectral diffusion (black dashed curve in Figure~\ref{fig:Fig_IV}a). 
The interference visibility was measured to be $ V_{\mathrm{HOM}}(0) = 0.80 \pm 0.06 $ for the resonant case without any postprocessing. 
For the moderately detuned case, the tunable optical transitions were shifted to introduce a 
frequency detuning of \SI{800}{\mega\hertz} between photons. The resulting dip, presented in 
Figure~\ref{fig:Fig_IV}b, shows interference between single photons yielding 
$ V_{\mathrm{HOM}}(0) = 0.63 \pm 0.08 $.

The maximally detuned case was measured with the setpoint at the maximum allowed 
voltage of 130~V that did not introduce heating of the cryostat. The resulting autocorrelation 
function in Figure~\ref{fig:Fig_IV}c shows that at maximum detuning, two-photon interference 
signatures are no longer visible within our experimental detection time resolution -- a result of combined large detuning and spectral diffusion 
measured at high voltage~\cite{morioka_spin-controlled_2020, rezai_coherence_2018}.

To model the measured second-order autocorrelation function in the case of two-photon interference~\cite{legero_quantum_2004, lettow_quantum_2010}, 
we used a theoretical framework that accounts for the coherence properties of the emitters and 
the effects of spectral diffusion (see Methods for details). This model (equation~\ref{eq:g2_sd}) was used to produce the red curve, which fits the 
experimental data in Figure~\ref{fig:Fig_IV}.

\section{Discussion}\label{sec:discussion}
In this study, we demonstrate the successful generation of 
SnV\textsuperscript{--} defects with a narrow inhomogeneous distribution without the use of HPHT annealing~\cite{narita_multiple_2023}. 
Although HPHT annealing effectively reduces 
inhomogeneous broadening, it can significantly degrade surface quality, making it less compatible 
with nanofabrication. By employing channeled implantation, we achieved a narrow 
inhomogeneous distribution for SnV\textsuperscript{--} centers. 
Defects in these samples exhibited highly coherent optical transitions 
and long-term spectral stability, even under off-resonant excitation. In the absence of internal stress 
within the diamond lattice, caused by lattice damage~\cite{aghaeimeibodi_electrical_2021}, tin vacancies remain unpolarized, 
making them less susceptible to electric field fluctuations~\cite{li_atomic_2024,pieplow_quantum_2024}.

Furthermore, the obtained narrow inhomogeneous distribution was compensated for by Stark shift tuning, which
preserved optical coherence and photon emission quality. This tuning approach, 
unlike strain-based methods~\cite{machielse_quantum_2019, wan_large-scale_2020}, allows 
for precise emission frequency alignment with minimal induced dipole moment in the defect, 
maintaining spectral stability.

We demonstrated two-photon interference experiments between negatively 
charged tin-vacancy (SnV\textsuperscript{--}) centers in separate quantum nodes, 
achieving a visibility of $80 \pm 6\%$ at zero detuning. The measured high visibility illustrates 
the high optical coherence of SnV\textsuperscript{--} centers despite off-resonant excitation. 
The observed spectral diffusion aligns with the presence of a nearby charge trap or multiple traps, as evidenced by 
the fitted kink in 
the presented Stark shift measurements in Figure~\ref{fig:Fig_II}c. These findings suggest 
that the off-resonant excitation approach for single-photon generation is suboptimal, 
since it induces population fluctuations of the charge traps in the defect's vicinity.
Resonant excitation is preferable, since it would selectively excite the SnV\textsuperscript{--}, 
yielding Fourier-limited emission, thereby maximizing the interference 
probability~\cite{kambs_limitations_2018, rezai_coherence_2018}.

In our detuned interference measurements, we observe that two-photon interference 
remains feasible even when the photons are not in resonance. The interference in the detuned case is possible because the coherence time 
of the emitted photons is sufficiently long to exceed the detection time resolution 
\cite{pittman_can_1996, legero_time-resolved_2003, legero_quantum_2004, halder_entangling_2007}. 
Notably, the interference occurs at the detection stage rather than at the beam splitter, 
with the measurement time bin acting as a temporal filter. This filter broadens 
the observed frequency resolution to $\frac{2\pi}{\Delta \tau}$, effectively "erasing" the which-frequency information 
\cite{wheeler_past_1978}. Furthermore, such detuned interference can be useful for 
entangling photons of different colors~\cite{zhao_entangling_2014}, enabling the creation of entangled states 
between non-identical quantum systems~\cite{parker_diamond_2024}.
In conclusion, this study establishes SnV\textsuperscript{--} centers as a scalable, 
promising platform for quantum networking. 
The generation of SnV\textsuperscript{--} centers with narrow inhomogeneous distribution,
combined with Stark tuning and robust two-photon 
interference, positions these centers as key candidates for large-scale quantum applications.
\begin{acknowledgments}
We acknowledge fruitful discussions with Florian Kaiser, Jonathan Körber, Rainer Stöhr, Javid Javadzade, and Roman Kolesov. 
This work was funded by the Spinning project (BMBF, No.\ 13N16219), QR.X project (BMBF, No.\ 16KISQ013), QMinT (BMBF No.\ 13N15976),
by the European Union via SPINUS, by the German Research Foundation (DFG, No.\ RTG 2642), and 
the state of Baden-W\"urttemberg (QC4BW and KQCBW24).
O.B. acknowledges support from the RTG 2642.
\end{acknowledgments}
\section*{Appendices}

\subsection{Experimental setup}
The two-node setup includes two cryostats, the AttoDry800 (node I) and Bluefors LD250 (node II) (see Figure~\ref{fig:Fig_III}a), 
located in separate rooms and connected by a 70~m long polarization-maintaining (PM) fiber. 
The setup directs the green excitation beam through a 567~nm long-pass filter into the cryostat, where it is focused by a 0.9~NA, 100× objective (Olympus MPLN). 
Emission from the SnV\textsuperscript{--} is then separated by a 630~nm long-pass filter, 
guiding the phonon sideband and coherent zero-phonon line emission to the avalanche photodiodes (APDs) (SPCM-AQRH-14-FC, Excelitas Technologies Corp.). 
The zero-phonon lines are collected via a 70~m PM fiber for node I and a 5~m PM fiber for node II, and then coupled into an interferometer.
Inside the interferometer, emission from both nodes is combined into a single spatial mode with orthogonal polarizations 
using half-wave plates and a polarizing beam splitter (PBS I). This mode is further filtered by a tilted 0.5~nm narrow-band-pass filter (LC-HBP630.2/0.5, Alluxa, Inc.), 
isolating the desired line. Due to the substantial spectral separation between emission lines stemming from large spin-orbit coupling, 
a narrow etalon is not required, which allows for coherent transition selection with a commercial spectral filter.
The combined and filtered emission is split again by a second polarizing beam splitter. This configuration allows two modes of operation: (1) 
with a half-wave plate before PBS II maintaining separate polarizations, enabling independent measurements from nodes I and II; and (2) 
with the half-wave plate rotated by $45^\circ$, mixing polarizations to observe interference~\cite{beugnon_quantum_2006}. 

The measured photon arrival times are analyzed using time-stamping electronics (Timetagger 20, Swabian Instruments), allowing precise time-correlation measurements.
Additionally, a resonant laser (620~$\pm$~10~nm DLC TA-SHG PRO, TOPTICA Photonics AG) is introduced into the setup through PBS I, coupling into the PM fibers of both nodes.
This design allows the collected zero-phonon line emission (from green excitation) to share the spatial mode with the resonant laser, enabling 
clear identification of defects under both off-resonant and resonant excitation conditions.
The remote control of both nodes was performed using Qudi software
with customized modules~\cite{binder_qudi_2017} . 

\subsection{Sample fabrication}
In this work, we implanted tin ions  into [001]-oriented CVD-grown diamond at energies
 of \SI{40}{\kilo\electronvolt}, \SI{80}{\kilo\electronvolt}, and \SI{170}{\kilo\electronvolt} at a $0^\circ$ incident angle and \SI{5e10}{\centi\meter^{-2}}. 
Following implantation, high-temperature annealing at \SI{1500}{\celsius} was performed for 1 hour. 
As a final step, 3-acid boiling was employed for 3 hours to remove the graphitized surface layer.

\subsection{Electrodes fabrication}
The electrodes are spaced \SI{1.7}{\micro\meter} apart and are \SI{1.8}{\micro\meter} wide. 
 They were created using a liftoff process, with the electrode shape patterned into a double layer of photoresist (LOR3A and Shipley S1805) via optical lithography. 
 The shape begins as two narrow, parallel electrodes that gradually widen and separate until they terminate in wire-bonding pads. The patterned 
 shape is slightly narrower and farther apart than the final electrodes. 
 These adjustments help to prevent the electrodes from fusing due to the proximity effect.
 Following the patterning, \SI{10}{\nano\meter} of Cr as a mating layer, followed by \SI{100}{\nano\meter} 
 of Au, were deposited onto the sample by e-beam evaporation at a rate of \SI{1}{\angstrom\per\second}.
 The remaining photoresist was dissolved in N-Methyl-2-pyrrolidone (NMP), which lifts off the unwanted 
 gold, leaving just the electrodes depicted in Figure~\ref{fig:Fig_III}a. 

\subsection{Modeling the Stark shift}
To model the observed kink in the Stark shift curve 
caused by a nearby charge trap, we consider the following 
higher-order polynomial terms~\cite{de_santis_investigation_2021} in the spectral shift of the defect:

\begin{align}
  h\Delta \nu_{\pm}(\mathcal{E}) &= -\mu_{\mathrm{tin}} (\mathcal{E} \pm \mathcal{E}_{\mathrm{trap}}) 
  - \frac{\alpha}{2}(\mathcal{E} \pm \mathcal{E}_{\mathrm{trap}})^2 \nonumber \\
  &\quad - \frac{\beta}{6}(\mathcal{E} \pm \mathcal{E}_{\mathrm{trap}})^3 
  - \frac{\gamma}{24}(\mathcal{E} \pm \mathcal{E}_{\mathrm{trap}})^4,
\end{align}

where $\mathcal{E}$ is the applied voltage combined with the static internal electric field (\textit{i.e.}, strain or surface-induced), $\alpha$, $\beta$, and $\gamma$ 
represent the second-, third-, and fourth-order Stark coefficients, respectively, and $\mu$ is the linear Stark coefficient. 
The term $\mathcal{E}_{\mathrm{trap}}$ denotes the electric field offset from the charge trap in the defect's vicinity.

To account for the probability $p$ of the trap flipping in response to the applied field, we follow the definition by Segura et al.\cite{segura_tip-induced_2001}:

\begin{align} 
  p &= 1 / \left[1 + \exp\left(\frac{A_0 + 2\mu_{\mathrm{trap}} \mathcal{E}}{k_{\mathrm{B}} T}\right)\right], 
\end{align}

where $\mu_{\mathrm{trap}}$ is the charge trap's dipole moment, $A_0$ represents the external field dependence of the charge trap population.

The combined spectral shift, considering both filled and empty states 
of the trap, is then expressed as:

\begin{align}
  h\Delta \nu_{\text{total}}(\mathcal{E}) &= p \cdot  h\Delta \nu_{+}(\mathcal{E}) + (1 - p) \cdot h\Delta \nu_{-}(\mathcal{E}).
\end{align}

Here, $ h\Delta \nu_{+}(\mathcal{E}) $ and $ h\Delta \nu_{-}(\mathcal{E}) $ represent the Stark shifts 
when the trap is in the filled and empty states, respectively. 
The resulting spectral shift $ h\Delta \nu_{\text{total}}(\mathcal{E}) $ captures the 
overall behavior observed in the experimental data (black curve fit in Figure~\ref{fig:Fig_II}c), showing a kink at 
the point where the trap transitions between states.

\subsection{Modeling of two-photon interference}
To model the measured second-order cross-correlation function in the case of two-photon interference, 
we use the expression for $ g^{(2)}_{\mathrm{tpi}}(\tau) $ as described in~\cite{lettow_quantum_2010}:

\begin{multline} \label{eq:g2_tpi}
  g^{(2)}_{\mathrm{tpi}}(\tau) = c_1^2 g^{(2)}_{11}(\tau) + c_2^2 g^{(2)}_{22}(\tau) \\
  + 2c_1 c_2 \left\{ 1 - \eta \frac{ S_1 S_2 }{I_1 I_2 }\left|g^{(1)}_{11}(\tau)\right| \left|g^{(1)}_{22}(\tau)\right| \cos(\Delta \omega \tau) \right\},
\end{multline}

where $ c_i = \frac{I_i}{I_1 + I_2} $ represents the normalized intensity ratio of source $ i $, 
$ g^{(2)}_{ii}(\tau) $ is the second-order autocorrelation function of source $ i $,
$ g^{(1)}_{ii}(\tau) $ is the first-order autocorrelation function of source $ i $,
$ \eta $ is a visibility reduction factor accounting for experimental factors such as classical interferometer visibility,
$ S_i $ is the signal count rate from source $ i $,
$  I_i  $ is the total detected count rate from source $ i $, including background,
$ \Delta \omega = \omega_1 - \omega_2 $ is the frequency detuning between the two sources,
$ \tau $ is the time delay between photon detection events.

The first two terms in equation~\ref{eq:g2_tpi} represent the second-order autocorrelations of each individual source. 
The third term accounts for the interference between the two sources, depending on their coherence properties and the frequency detuning $ \Delta \omega $.

For a quantum emitter under continuous excitation, the magnitude of the first-order autocorrelation function describes the 
decay of coherence over time due to dephasing processes:

\begin{equation} \label{eq:g1_decay}
  \left|g^{(1)}_{ii}(\tau) \right| = e^{ - \gamma \tau / 2 },
\end{equation}

where $ \gamma = \frac{1}{\tau_{\mathrm{coh}}} $ is the homogeneous linewidth of the emitter, and $ \tau_{\mathrm{coh}} $ 
is the coherence time of the emitter. The extent of coherence of the source can be estimated by how closely the emitter linewidth approaches 
the Fourier limit. In the case of transform-limited lines, the coherence time reaches its maximum value, $ \tau_{\mathrm{coh}} = 2 \tau_{\mathrm{rad}} $, 
which is twice the radiative lifetime~\cite{wang_optical_2022}.

In the presence of spectral diffusion, the beating signal described by equation~\ref{eq:g2_tpi} becomes 
washed out due to contributions from multiple beating frequencies. To account for spectral diffusion, 
we modify the function accordingly. Assuming a Gaussian distribution for the detuning variations, 
the modified function can be expressed as a convolution:

\begin{equation}\label{eq:g2_sd}
  g^{(2)}_{\mathrm{sd}}(\tau) = \int_{-\infty}^{\infty} \mathcal{G}_{\sigma}(\Delta \omega') \, g^{(2)}_{\mathrm{tpi}}(\tau, \Delta \omega') \, d\Delta \omega',
\end{equation}

where $\mathcal{G}_{\sigma}(\Delta \omega')$
is the Gaussian kernel characterizing the spectral diffusion, $\sigma$ is the
standard deviation of the frequency distribution caused by spectral diffusion, and $\omega_0$ is the central frequency.

\bibliography{TPI}

\begin{thebibliography}{69}%
\makeatletter
\providecommand \@ifxundefined [1]{%
 \@ifx{#1\undefined}
}%
\providecommand \@ifnum [1]{%
 \ifnum #1\expandafter \@firstoftwo
 \else \expandafter \@secondoftwo
 \fi
}%
\providecommand \@ifx [1]{%
 \ifx #1\expandafter \@firstoftwo
 \else \expandafter \@secondoftwo
 \fi
}%
\providecommand \natexlab [1]{#1}%
\providecommand \enquote  [1]{``#1''}%
\providecommand \bibnamefont  [1]{#1}%
\providecommand \bibfnamefont [1]{#1}%
\providecommand \citenamefont [1]{#1}%
\providecommand \href@noop [0]{\@secondoftwo}%
\providecommand \href [0]{\begingroup \@sanitize@url \@href}%
\providecommand \@href[1]{\@@startlink{#1}\@@href}%
\providecommand \@@href[1]{\endgroup#1\@@endlink}%
\providecommand \@sanitize@url [0]{\catcode `\\12\catcode `\$12\catcode `\&12\catcode `\#12\catcode `\^12\catcode `\_12\catcode `\%12\relax}%
\providecommand \@@startlink[1]{}%
\providecommand \@@endlink[0]{}%
\providecommand \url  [0]{\begingroup\@sanitize@url \@url }%
\providecommand \@url [1]{\endgroup\@href {#1}{\urlprefix }}%
\providecommand \urlprefix  [0]{URL }%
\providecommand \Eprint [0]{\href }%
\providecommand \doibase [0]{https://doi.org/}%
\providecommand \selectlanguage [0]{\@gobble}%
\providecommand \bibinfo  [0]{\@secondoftwo}%
\providecommand \bibfield  [0]{\@secondoftwo}%
\providecommand \translation [1]{[#1]}%
\providecommand \BibitemOpen [0]{}%
\providecommand \bibitemStop [0]{}%
\providecommand \bibitemNoStop [0]{.\EOS\space}%
\providecommand \EOS [0]{\spacefactor3000\relax}%
\providecommand \BibitemShut  [1]{\csname bibitem#1\endcsname}%
\let\auto@bib@innerbib\@empty
\bibitem [{\citenamefont {Awschalom}\ \emph {et~al.}(2021)\citenamefont {Awschalom}, \citenamefont {Berggren}, \citenamefont {Bernien}, \citenamefont {Bhave}, \citenamefont {Carr}, \citenamefont {Davids}, \citenamefont {Economou}, \citenamefont {Englund}, \citenamefont {Faraon}, \citenamefont {Fejer}, \citenamefont {Guha}, \citenamefont {Gustafsson}, \citenamefont {Hu}, \citenamefont {Jiang}, \citenamefont {Kim}, \citenamefont {Korzh}, \citenamefont {Kumar}, \citenamefont {Kwiat}, \citenamefont {Lon{\v c}ar}, \citenamefont {Lukin}, \citenamefont {Miller}, \citenamefont {Monroe}, \citenamefont {Nam}, \citenamefont {Narang}, \citenamefont {Orcutt}, \citenamefont {Raymer}, \citenamefont {Safavi-Naeini}, \citenamefont {Spiropulu}, \citenamefont {Srinivasan}, \citenamefont {Sun}, \citenamefont {Vu{\v c}kovi{\'c}}, \citenamefont {Waks}, \citenamefont {Walsworth}, \citenamefont {Weiner},\ and\ \citenamefont {Zhang}}]{awschalom_development_2021}%
  \BibitemOpen
  \bibfield  {author} {\bibinfo {author} {\bibfnamefont {D.}~\bibnamefont {Awschalom}}, \bibinfo {author} {\bibfnamefont {K.~K.}\ \bibnamefont {Berggren}}, \bibinfo {author} {\bibfnamefont {H.}~\bibnamefont {Bernien}}, \bibinfo {author} {\bibfnamefont {S.}~\bibnamefont {Bhave}}, \bibinfo {author} {\bibfnamefont {L.~D.}\ \bibnamefont {Carr}}, \bibinfo {author} {\bibfnamefont {P.}~\bibnamefont {Davids}}, \bibinfo {author} {\bibfnamefont {S.~E.}\ \bibnamefont {Economou}}, \bibinfo {author} {\bibfnamefont {D.}~\bibnamefont {Englund}}, \bibinfo {author} {\bibfnamefont {A.}~\bibnamefont {Faraon}}, \bibinfo {author} {\bibfnamefont {M.}~\bibnamefont {Fejer}}, \bibinfo {author} {\bibfnamefont {S.}~\bibnamefont {Guha}}, \bibinfo {author} {\bibfnamefont {M.~V.}\ \bibnamefont {Gustafsson}}, \bibinfo {author} {\bibfnamefont {E.}~\bibnamefont {Hu}}, \bibinfo {author} {\bibfnamefont {L.}~\bibnamefont {Jiang}}, \bibinfo {author} {\bibfnamefont {J.}~\bibnamefont {Kim}}, \bibinfo {author} {\bibfnamefont {B.}~\bibnamefont
  {Korzh}}, \bibinfo {author} {\bibfnamefont {P.}~\bibnamefont {Kumar}}, \bibinfo {author} {\bibfnamefont {P.~G.}\ \bibnamefont {Kwiat}}, \bibinfo {author} {\bibfnamefont {M.}~\bibnamefont {Lon{\v c}ar}}, \bibinfo {author} {\bibfnamefont {M.~D.}\ \bibnamefont {Lukin}}, \bibinfo {author} {\bibfnamefont {D.~A.}\ \bibnamefont {Miller}}, \bibinfo {author} {\bibfnamefont {C.}~\bibnamefont {Monroe}}, \bibinfo {author} {\bibfnamefont {S.~W.}\ \bibnamefont {Nam}}, \bibinfo {author} {\bibfnamefont {P.}~\bibnamefont {Narang}}, \bibinfo {author} {\bibfnamefont {J.~S.}\ \bibnamefont {Orcutt}}, \bibinfo {author} {\bibfnamefont {M.~G.}\ \bibnamefont {Raymer}}, \bibinfo {author} {\bibfnamefont {A.~H.}\ \bibnamefont {Safavi-Naeini}}, \bibinfo {author} {\bibfnamefont {M.}~\bibnamefont {Spiropulu}}, \bibinfo {author} {\bibfnamefont {K.}~\bibnamefont {Srinivasan}}, \bibinfo {author} {\bibfnamefont {S.}~\bibnamefont {Sun}}, \bibinfo {author} {\bibfnamefont {J.}~\bibnamefont {Vu{\v c}kovi{\'c}}}, \bibinfo {author} {\bibfnamefont
  {E.}~\bibnamefont {Waks}}, \bibinfo {author} {\bibfnamefont {R.}~\bibnamefont {Walsworth}}, \bibinfo {author} {\bibfnamefont {A.~M.}\ \bibnamefont {Weiner}},\ and\ \bibinfo {author} {\bibfnamefont {Z.}~\bibnamefont {Zhang}},\ }\bibfield  {title} {\bibinfo {title} {Development of {Quantum} {Interconnects} ({QuICs}) for {Next}-{Generation} {Information} {Technologies}},\ }\href {https://doi.org/10.1103/PRXQuantum.2.017002} {\bibfield  {journal} {\bibinfo  {journal} {PRX Quantum}\ }\textbf {\bibinfo {volume} {2}},\ \bibinfo {pages} {017002} (\bibinfo {year} {2021})},\ \bibinfo {note} {publisher: American Physical Society}\BibitemShut {NoStop}%
\bibitem [{\citenamefont {Ritter}\ \emph {et~al.}(2012)\citenamefont {Ritter}, \citenamefont {N{\"o}lleke}, \citenamefont {Hahn}, \citenamefont {Reiserer}, \citenamefont {Neuzner}, \citenamefont {Uphoff}, \citenamefont {M{\"u}cke}, \citenamefont {Figueroa}, \citenamefont {Bochmann},\ and\ \citenamefont {Rempe}}]{ritter_elementary_2012}%
  \BibitemOpen
  \bibfield  {author} {\bibinfo {author} {\bibfnamefont {S.}~\bibnamefont {Ritter}}, \bibinfo {author} {\bibfnamefont {C.}~\bibnamefont {N{\"o}lleke}}, \bibinfo {author} {\bibfnamefont {C.}~\bibnamefont {Hahn}}, \bibinfo {author} {\bibfnamefont {A.}~\bibnamefont {Reiserer}}, \bibinfo {author} {\bibfnamefont {A.}~\bibnamefont {Neuzner}}, \bibinfo {author} {\bibfnamefont {M.}~\bibnamefont {Uphoff}}, \bibinfo {author} {\bibfnamefont {M.}~\bibnamefont {M{\"u}cke}}, \bibinfo {author} {\bibfnamefont {E.}~\bibnamefont {Figueroa}}, \bibinfo {author} {\bibfnamefont {J.}~\bibnamefont {Bochmann}},\ and\ \bibinfo {author} {\bibfnamefont {G.}~\bibnamefont {Rempe}},\ }\bibfield  {title} {\bibinfo {title} {An elementary quantum network of single atoms in optical cavities},\ }\href {https://doi.org/10.1038/nature11023} {\bibfield  {journal} {\bibinfo  {journal} {Nature}\ }\textbf {\bibinfo {volume} {484}},\ \bibinfo {pages} {195} (\bibinfo {year} {2012})}\BibitemShut {NoStop}%
\bibitem [{\citenamefont {Hofmann}\ \emph {et~al.}(2012)\citenamefont {Hofmann}, \citenamefont {Krug}, \citenamefont {Ortegel}, \citenamefont {G{\'e}rard}, \citenamefont {Weber}, \citenamefont {Rosenfeld},\ and\ \citenamefont {Weinfurter}}]{hofmann_heralded_2012}%
  \BibitemOpen
  \bibfield  {author} {\bibinfo {author} {\bibfnamefont {J.}~\bibnamefont {Hofmann}}, \bibinfo {author} {\bibfnamefont {M.}~\bibnamefont {Krug}}, \bibinfo {author} {\bibfnamefont {N.}~\bibnamefont {Ortegel}}, \bibinfo {author} {\bibfnamefont {L.}~\bibnamefont {G{\'e}rard}}, \bibinfo {author} {\bibfnamefont {M.}~\bibnamefont {Weber}}, \bibinfo {author} {\bibfnamefont {W.}~\bibnamefont {Rosenfeld}},\ and\ \bibinfo {author} {\bibfnamefont {H.}~\bibnamefont {Weinfurter}},\ }\bibfield  {title} {\bibinfo {title} {Heralded {Entanglement} {Between} {Widely} {Separated} {Atoms}},\ }\href {https://doi.org/10.1126/science.1221856} {\bibfield  {journal} {\bibinfo  {journal} {Science}\ }\textbf {\bibinfo {volume} {337}},\ \bibinfo {pages} {72} (\bibinfo {year} {2012})}\BibitemShut {NoStop}%
\bibitem [{\citenamefont {Bock}\ \emph {et~al.}(2018)\citenamefont {Bock}, \citenamefont {Eich}, \citenamefont {Kucera}, \citenamefont {Kreis}, \citenamefont {Lenhard}, \citenamefont {Becher},\ and\ \citenamefont {Eschner}}]{bock_high-fidelity_2018}%
  \BibitemOpen
  \bibfield  {author} {\bibinfo {author} {\bibfnamefont {M.}~\bibnamefont {Bock}}, \bibinfo {author} {\bibfnamefont {P.}~\bibnamefont {Eich}}, \bibinfo {author} {\bibfnamefont {S.}~\bibnamefont {Kucera}}, \bibinfo {author} {\bibfnamefont {M.}~\bibnamefont {Kreis}}, \bibinfo {author} {\bibfnamefont {A.}~\bibnamefont {Lenhard}}, \bibinfo {author} {\bibfnamefont {C.}~\bibnamefont {Becher}},\ and\ \bibinfo {author} {\bibfnamefont {J.}~\bibnamefont {Eschner}},\ }\bibfield  {title} {\bibinfo {title} {High-fidelity entanglement between a trapped ion and a telecom photon via quantum frequency conversion},\ }\href {https://doi.org/10.1038/s41467-018-04341-2} {\bibfield  {journal} {\bibinfo  {journal} {Nature Communications}\ }\textbf {\bibinfo {volume} {9}},\ \bibinfo {pages} {1998} (\bibinfo {year} {2018})}\BibitemShut {NoStop}%
\bibitem [{\citenamefont {Reiserer}\ \emph {et~al.}(2014)\citenamefont {Reiserer}, \citenamefont {Kalb}, \citenamefont {Rempe},\ and\ \citenamefont {Ritter}}]{reiserer_quantum_2014}%
  \BibitemOpen
  \bibfield  {author} {\bibinfo {author} {\bibfnamefont {A.}~\bibnamefont {Reiserer}}, \bibinfo {author} {\bibfnamefont {N.}~\bibnamefont {Kalb}}, \bibinfo {author} {\bibfnamefont {G.}~\bibnamefont {Rempe}},\ and\ \bibinfo {author} {\bibfnamefont {S.}~\bibnamefont {Ritter}},\ }\bibfield  {title} {\bibinfo {title} {A quantum gate between a flying optical photon and a single trapped atom},\ }\href {https://doi.org/10.1038/nature13177} {\bibfield  {journal} {\bibinfo  {journal} {Nature}\ }\textbf {\bibinfo {volume} {508}},\ \bibinfo {pages} {237} (\bibinfo {year} {2014})},\ \bibinfo {note} {publisher: Nature Publishing Group}\BibitemShut {NoStop}%
\bibitem [{\citenamefont {Hucul}\ \emph {et~al.}(2015)\citenamefont {Hucul}, \citenamefont {Inlek}, \citenamefont {Vittorini}, \citenamefont {Crocker}, \citenamefont {Debnath}, \citenamefont {Clark},\ and\ \citenamefont {Monroe}}]{hucul_modular_2015}%
  \BibitemOpen
  \bibfield  {author} {\bibinfo {author} {\bibfnamefont {D.}~\bibnamefont {Hucul}}, \bibinfo {author} {\bibfnamefont {I.~V.}\ \bibnamefont {Inlek}}, \bibinfo {author} {\bibfnamefont {G.}~\bibnamefont {Vittorini}}, \bibinfo {author} {\bibfnamefont {C.}~\bibnamefont {Crocker}}, \bibinfo {author} {\bibfnamefont {S.}~\bibnamefont {Debnath}}, \bibinfo {author} {\bibfnamefont {S.~M.}\ \bibnamefont {Clark}},\ and\ \bibinfo {author} {\bibfnamefont {C.}~\bibnamefont {Monroe}},\ }\bibfield  {title} {\bibinfo {title} {Modular entanglement of atomic qubits using photons and phonons},\ }\href {https://doi.org/10.1038/nphys3150} {\bibfield  {journal} {\bibinfo  {journal} {Nature Physics}\ }\textbf {\bibinfo {volume} {11}},\ \bibinfo {pages} {37} (\bibinfo {year} {2015})},\ \bibinfo {note} {publisher: Nature Publishing Group}\BibitemShut {NoStop}%
\bibitem [{\citenamefont {Rosenfeld}\ \emph {et~al.}(2017)\citenamefont {Rosenfeld}, \citenamefont {Burchardt}, \citenamefont {Garthoff}, \citenamefont {Redeker}, \citenamefont {Ortegel}, \citenamefont {Rau},\ and\ \citenamefont {Weinfurter}}]{rosenfeld_event-ready_2017}%
  \BibitemOpen
  \bibfield  {author} {\bibinfo {author} {\bibfnamefont {W.}~\bibnamefont {Rosenfeld}}, \bibinfo {author} {\bibfnamefont {D.}~\bibnamefont {Burchardt}}, \bibinfo {author} {\bibfnamefont {R.}~\bibnamefont {Garthoff}}, \bibinfo {author} {\bibfnamefont {K.}~\bibnamefont {Redeker}}, \bibinfo {author} {\bibfnamefont {N.}~\bibnamefont {Ortegel}}, \bibinfo {author} {\bibfnamefont {M.}~\bibnamefont {Rau}},\ and\ \bibinfo {author} {\bibfnamefont {H.}~\bibnamefont {Weinfurter}},\ }\bibfield  {title} {\bibinfo {title} {Event-{Ready} {Bell} {Test} {Using} {Entangled} {Atoms} {Simultaneously} {Closing} {Detection} and {Locality} {Loopholes}},\ }\href {https://doi.org/10.1103/PhysRevLett.119.010402} {\bibfield  {journal} {\bibinfo  {journal} {Physical Review Letters}\ }\textbf {\bibinfo {volume} {119}},\ \bibinfo {pages} {010402} (\bibinfo {year} {2017})}\BibitemShut {NoStop}%
\bibitem [{\citenamefont {Zaporski}\ \emph {et~al.}(2023)\citenamefont {Zaporski}, \citenamefont {Shofer}, \citenamefont {Bodey}, \citenamefont {Manna}, \citenamefont {Gillard}, \citenamefont {Appel}, \citenamefont {Schimpf}, \citenamefont {Covre Da~Silva}, \citenamefont {Jarman}, \citenamefont {Delamare}, \citenamefont {Park}, \citenamefont {Haeusler}, \citenamefont {Chekhovich}, \citenamefont {Rastelli}, \citenamefont {Gangloff}, \citenamefont {Atat{\"u}re},\ and\ \citenamefont {Le~Gall}}]{zaporski_ideal_2023}%
  \BibitemOpen
  \bibfield  {author} {\bibinfo {author} {\bibfnamefont {L.}~\bibnamefont {Zaporski}}, \bibinfo {author} {\bibfnamefont {N.}~\bibnamefont {Shofer}}, \bibinfo {author} {\bibfnamefont {J.~H.}\ \bibnamefont {Bodey}}, \bibinfo {author} {\bibfnamefont {S.}~\bibnamefont {Manna}}, \bibinfo {author} {\bibfnamefont {G.}~\bibnamefont {Gillard}}, \bibinfo {author} {\bibfnamefont {M.~H.}\ \bibnamefont {Appel}}, \bibinfo {author} {\bibfnamefont {C.}~\bibnamefont {Schimpf}}, \bibinfo {author} {\bibfnamefont {S.~F.}\ \bibnamefont {Covre Da~Silva}}, \bibinfo {author} {\bibfnamefont {J.}~\bibnamefont {Jarman}}, \bibinfo {author} {\bibfnamefont {G.}~\bibnamefont {Delamare}}, \bibinfo {author} {\bibfnamefont {G.}~\bibnamefont {Park}}, \bibinfo {author} {\bibfnamefont {U.}~\bibnamefont {Haeusler}}, \bibinfo {author} {\bibfnamefont {E.~A.}\ \bibnamefont {Chekhovich}}, \bibinfo {author} {\bibfnamefont {A.}~\bibnamefont {Rastelli}}, \bibinfo {author} {\bibfnamefont {D.~A.}\ \bibnamefont {Gangloff}}, \bibinfo {author} {\bibfnamefont
  {M.}~\bibnamefont {Atat{\"u}re}},\ and\ \bibinfo {author} {\bibfnamefont {C.}~\bibnamefont {Le~Gall}},\ }\bibfield  {title} {\bibinfo {title} {Ideal refocusing of an optically active spin qubit under strong hyperfine interactions},\ }\href {https://doi.org/10.1038/s41565-022-01282-2} {\bibfield  {journal} {\bibinfo  {journal} {Nature Nanotechnology}\ }\textbf {\bibinfo {volume} {18}},\ \bibinfo {pages} {257} (\bibinfo {year} {2023})}\BibitemShut {NoStop}%
\bibitem [{\citenamefont {Yu}\ \emph {et~al.}(2023)\citenamefont {Yu}, \citenamefont {Liu}, \citenamefont {Lee}, \citenamefont {Michler}, \citenamefont {Reitzenstein}, \citenamefont {Srinivasan}, \citenamefont {Waks},\ and\ \citenamefont {Liu}}]{yu_telecom-band_2023}%
  \BibitemOpen
  \bibfield  {author} {\bibinfo {author} {\bibfnamefont {Y.}~\bibnamefont {Yu}}, \bibinfo {author} {\bibfnamefont {S.}~\bibnamefont {Liu}}, \bibinfo {author} {\bibfnamefont {C.-M.}\ \bibnamefont {Lee}}, \bibinfo {author} {\bibfnamefont {P.}~\bibnamefont {Michler}}, \bibinfo {author} {\bibfnamefont {S.}~\bibnamefont {Reitzenstein}}, \bibinfo {author} {\bibfnamefont {K.}~\bibnamefont {Srinivasan}}, \bibinfo {author} {\bibfnamefont {E.}~\bibnamefont {Waks}},\ and\ \bibinfo {author} {\bibfnamefont {J.}~\bibnamefont {Liu}},\ }\bibfield  {title} {\bibinfo {title} {Telecom-band quantum dot technologies for long-distance quantum networks},\ }\href {https://doi.org/10.1038/s41565-023-01528-7} {\bibfield  {journal} {\bibinfo  {journal} {Nature Nanotechnology}\ }\textbf {\bibinfo {volume} {18}},\ \bibinfo {pages} {1389} (\bibinfo {year} {2023})}\BibitemShut {NoStop}%
\bibitem [{\citenamefont {Siyushev}\ \emph {et~al.}(2014)\citenamefont {Siyushev}, \citenamefont {Stein}, \citenamefont {Wrachtrup},\ and\ \citenamefont {Gerhardt}}]{siyushev_molecular_2014}%
  \BibitemOpen
  \bibfield  {author} {\bibinfo {author} {\bibfnamefont {P.}~\bibnamefont {Siyushev}}, \bibinfo {author} {\bibfnamefont {G.}~\bibnamefont {Stein}}, \bibinfo {author} {\bibfnamefont {J.}~\bibnamefont {Wrachtrup}},\ and\ \bibinfo {author} {\bibfnamefont {I.}~\bibnamefont {Gerhardt}},\ }\bibfield  {title} {\bibinfo {title} {Molecular photons interfaced with alkali atoms},\ }\href {https://doi.org/10.1038/nature13191} {\bibfield  {journal} {\bibinfo  {journal} {Nature}\ }\textbf {\bibinfo {volume} {509}},\ \bibinfo {pages} {66} (\bibinfo {year} {2014})}\BibitemShut {NoStop}%
\bibitem [{\citenamefont {Lago-Rivera}\ \emph {et~al.}(2021)\citenamefont {Lago-Rivera}, \citenamefont {Grandi}, \citenamefont {Rakonjac}, \citenamefont {Seri},\ and\ \citenamefont {de~Riedmatten}}]{lago-rivera_telecom-heralded_2021}%
  \BibitemOpen
  \bibfield  {author} {\bibinfo {author} {\bibfnamefont {D.}~\bibnamefont {Lago-Rivera}}, \bibinfo {author} {\bibfnamefont {S.}~\bibnamefont {Grandi}}, \bibinfo {author} {\bibfnamefont {J.~V.}\ \bibnamefont {Rakonjac}}, \bibinfo {author} {\bibfnamefont {A.}~\bibnamefont {Seri}},\ and\ \bibinfo {author} {\bibfnamefont {H.}~\bibnamefont {de~Riedmatten}},\ }\bibfield  {title} {\bibinfo {title} {Telecom-heralded entanglement between multimode solid-state quantum memories},\ }\href {https://doi.org/10.1038/s41586-021-03481-8} {\bibfield  {journal} {\bibinfo  {journal} {Nature}\ }\textbf {\bibinfo {volume} {594}},\ \bibinfo {pages} {37} (\bibinfo {year} {2021})},\ \bibinfo {note} {publisher: Nature Publishing Group}\BibitemShut {NoStop}%
\bibitem [{\citenamefont {Ruskuc}\ \emph {et~al.}(2024)\citenamefont {Ruskuc}, \citenamefont {Wu}, \citenamefont {Green}, \citenamefont {Hermans}, \citenamefont {Choi},\ and\ \citenamefont {Faraon}}]{ruskuc2024scalable}%
  \BibitemOpen
  \bibfield  {author} {\bibinfo {author} {\bibfnamefont {A.}~\bibnamefont {Ruskuc}}, \bibinfo {author} {\bibfnamefont {C.-J.}\ \bibnamefont {Wu}}, \bibinfo {author} {\bibfnamefont {E.}~\bibnamefont {Green}}, \bibinfo {author} {\bibfnamefont {S.~L.}\ \bibnamefont {Hermans}}, \bibinfo {author} {\bibfnamefont {J.}~\bibnamefont {Choi}},\ and\ \bibinfo {author} {\bibfnamefont {A.}~\bibnamefont {Faraon}},\ }\bibfield  {title} {\bibinfo {title} {Scalable multipartite entanglement of remote rare-earth ion qubits},\ }\href@noop {} {\bibfield  {journal} {\bibinfo  {journal} {arXiv preprint arXiv:2402.16224}\ } (\bibinfo {year} {2024})}\BibitemShut {NoStop}%
\bibitem [{\citenamefont {Kurpiers}\ \emph {et~al.}(2018)\citenamefont {Kurpiers}, \citenamefont {Magnard}, \citenamefont {Walter}, \citenamefont {Royer}, \citenamefont {Pechal}, \citenamefont {Heinsoo}, \citenamefont {Salath{\'e}}, \citenamefont {Akin}, \citenamefont {Storz}, \citenamefont {Besse}, \citenamefont {Gasparinetti}, \citenamefont {Blais},\ and\ \citenamefont {Wallraff}}]{kurpiers_deterministic_2018}%
  \BibitemOpen
  \bibfield  {author} {\bibinfo {author} {\bibfnamefont {P.}~\bibnamefont {Kurpiers}}, \bibinfo {author} {\bibfnamefont {P.}~\bibnamefont {Magnard}}, \bibinfo {author} {\bibfnamefont {T.}~\bibnamefont {Walter}}, \bibinfo {author} {\bibfnamefont {B.}~\bibnamefont {Royer}}, \bibinfo {author} {\bibfnamefont {M.}~\bibnamefont {Pechal}}, \bibinfo {author} {\bibfnamefont {J.}~\bibnamefont {Heinsoo}}, \bibinfo {author} {\bibfnamefont {Y.}~\bibnamefont {Salath{\'e}}}, \bibinfo {author} {\bibfnamefont {A.}~\bibnamefont {Akin}}, \bibinfo {author} {\bibfnamefont {S.}~\bibnamefont {Storz}}, \bibinfo {author} {\bibfnamefont {J.-C.}\ \bibnamefont {Besse}}, \bibinfo {author} {\bibfnamefont {S.}~\bibnamefont {Gasparinetti}}, \bibinfo {author} {\bibfnamefont {A.}~\bibnamefont {Blais}},\ and\ \bibinfo {author} {\bibfnamefont {A.}~\bibnamefont {Wallraff}},\ }\bibfield  {title} {\bibinfo {title} {Deterministic quantum state transfer and remote entanglement using microwave photons},\ }\href
  {https://doi.org/10.1038/s41586-018-0195-y} {\bibfield  {journal} {\bibinfo  {journal} {Nature}\ }\textbf {\bibinfo {volume} {558}},\ \bibinfo {pages} {264} (\bibinfo {year} {2018})},\ \bibinfo {note} {publisher: Nature Publishing Group}\BibitemShut {NoStop}%
\bibitem [{\citenamefont {Hensen}\ \emph {et~al.}(2015)\citenamefont {Hensen}, \citenamefont {Bernien}, \citenamefont {Dr{\'e}au}, \citenamefont {Reiserer}, \citenamefont {Kalb}, \citenamefont {Blok}, \citenamefont {Ruitenberg}, \citenamefont {Vermeulen}, \citenamefont {Schouten}, \citenamefont {Abell{\'a}n}, \citenamefont {Amaya}, \citenamefont {Pruneri}, \citenamefont {Mitchell}, \citenamefont {Markham}, \citenamefont {Twitchen}, \citenamefont {Elkouss}, \citenamefont {Wehner}, \citenamefont {Taminiau},\ and\ \citenamefont {Hanson}}]{hensen_loophole-free_2015}%
  \BibitemOpen
  \bibfield  {author} {\bibinfo {author} {\bibfnamefont {B.}~\bibnamefont {Hensen}}, \bibinfo {author} {\bibfnamefont {H.}~\bibnamefont {Bernien}}, \bibinfo {author} {\bibfnamefont {A.~E.}\ \bibnamefont {Dr{\'e}au}}, \bibinfo {author} {\bibfnamefont {A.}~\bibnamefont {Reiserer}}, \bibinfo {author} {\bibfnamefont {N.}~\bibnamefont {Kalb}}, \bibinfo {author} {\bibfnamefont {M.~S.}\ \bibnamefont {Blok}}, \bibinfo {author} {\bibfnamefont {J.}~\bibnamefont {Ruitenberg}}, \bibinfo {author} {\bibfnamefont {R.~F.~L.}\ \bibnamefont {Vermeulen}}, \bibinfo {author} {\bibfnamefont {R.~N.}\ \bibnamefont {Schouten}}, \bibinfo {author} {\bibfnamefont {C.}~\bibnamefont {Abell{\'a}n}}, \bibinfo {author} {\bibfnamefont {W.}~\bibnamefont {Amaya}}, \bibinfo {author} {\bibfnamefont {V.}~\bibnamefont {Pruneri}}, \bibinfo {author} {\bibfnamefont {M.~W.}\ \bibnamefont {Mitchell}}, \bibinfo {author} {\bibfnamefont {M.}~\bibnamefont {Markham}}, \bibinfo {author} {\bibfnamefont {D.~J.}\ \bibnamefont {Twitchen}}, \bibinfo {author}
  {\bibfnamefont {D.}~\bibnamefont {Elkouss}}, \bibinfo {author} {\bibfnamefont {S.}~\bibnamefont {Wehner}}, \bibinfo {author} {\bibfnamefont {T.~H.}\ \bibnamefont {Taminiau}},\ and\ \bibinfo {author} {\bibfnamefont {R.}~\bibnamefont {Hanson}},\ }\bibfield  {title} {\bibinfo {title} {Loophole-free {Bell} inequality violation using electron spins separated by 1.3 kilometres},\ }\href {https://doi.org/10.1038/nature15759} {\bibfield  {journal} {\bibinfo  {journal} {Nature}\ }\textbf {\bibinfo {volume} {526}},\ \bibinfo {pages} {682} (\bibinfo {year} {2015})},\ \bibinfo {note} {publisher: Nature Publishing Group}\BibitemShut {NoStop}%
\bibitem [{\citenamefont {Awschalom}\ \emph {et~al.}(2018)\citenamefont {Awschalom}, \citenamefont {Hanson}, \citenamefont {Wrachtrup},\ and\ \citenamefont {Zhou}}]{awschalom2018quantum}%
  \BibitemOpen
  \bibfield  {author} {\bibinfo {author} {\bibfnamefont {D.~D.}\ \bibnamefont {Awschalom}}, \bibinfo {author} {\bibfnamefont {R.}~\bibnamefont {Hanson}}, \bibinfo {author} {\bibfnamefont {J.}~\bibnamefont {Wrachtrup}},\ and\ \bibinfo {author} {\bibfnamefont {B.~B.}\ \bibnamefont {Zhou}},\ }\bibfield  {title} {\bibinfo {title} {Quantum technologies with optically interfaced solid-state spins},\ }\href@noop {} {\bibfield  {journal} {\bibinfo  {journal} {Nature Photonics}\ }\textbf {\bibinfo {volume} {12}},\ \bibinfo {pages} {516} (\bibinfo {year} {2018})}\BibitemShut {NoStop}%
\bibitem [{\citenamefont {Higginbottom}\ \emph {et~al.}(2022)\citenamefont {Higginbottom}, \citenamefont {Kurkjian}, \citenamefont {Chartrand}, \citenamefont {Kazemi}, \citenamefont {Brunelle}, \citenamefont {MacQuarrie}, \citenamefont {Klein}, \citenamefont {Lee-Hone}, \citenamefont {Stacho}, \citenamefont {Ruether} \emph {et~al.}}]{higginbottom2022optical}%
  \BibitemOpen
  \bibfield  {author} {\bibinfo {author} {\bibfnamefont {D.~B.}\ \bibnamefont {Higginbottom}}, \bibinfo {author} {\bibfnamefont {A.~T.}\ \bibnamefont {Kurkjian}}, \bibinfo {author} {\bibfnamefont {C.}~\bibnamefont {Chartrand}}, \bibinfo {author} {\bibfnamefont {M.}~\bibnamefont {Kazemi}}, \bibinfo {author} {\bibfnamefont {N.~A.}\ \bibnamefont {Brunelle}}, \bibinfo {author} {\bibfnamefont {E.~R.}\ \bibnamefont {MacQuarrie}}, \bibinfo {author} {\bibfnamefont {J.~R.}\ \bibnamefont {Klein}}, \bibinfo {author} {\bibfnamefont {N.~R.}\ \bibnamefont {Lee-Hone}}, \bibinfo {author} {\bibfnamefont {J.}~\bibnamefont {Stacho}}, \bibinfo {author} {\bibfnamefont {M.}~\bibnamefont {Ruether}}, \emph {et~al.},\ }\bibfield  {title} {\bibinfo {title} {Optical observation of single spins in silicon},\ }\href@noop {} {\bibfield  {journal} {\bibinfo  {journal} {Nature}\ }\textbf {\bibinfo {volume} {607}},\ \bibinfo {pages} {266} (\bibinfo {year} {2022})}\BibitemShut {NoStop}%
\bibitem [{\citenamefont {Morioka}\ \emph {et~al.}(2020{\natexlab{a}})\citenamefont {Morioka}, \citenamefont {Babin}, \citenamefont {Nagy}, \citenamefont {Gediz}, \citenamefont {Hesselmeier}, \citenamefont {Liu}, \citenamefont {Joliffe}, \citenamefont {Niethammer}, \citenamefont {Dasari}, \citenamefont {Vorobyov} \emph {et~al.}}]{morioka2020spin}%
  \BibitemOpen
  \bibfield  {author} {\bibinfo {author} {\bibfnamefont {N.}~\bibnamefont {Morioka}}, \bibinfo {author} {\bibfnamefont {C.}~\bibnamefont {Babin}}, \bibinfo {author} {\bibfnamefont {R.}~\bibnamefont {Nagy}}, \bibinfo {author} {\bibfnamefont {I.}~\bibnamefont {Gediz}}, \bibinfo {author} {\bibfnamefont {E.}~\bibnamefont {Hesselmeier}}, \bibinfo {author} {\bibfnamefont {D.}~\bibnamefont {Liu}}, \bibinfo {author} {\bibfnamefont {M.}~\bibnamefont {Joliffe}}, \bibinfo {author} {\bibfnamefont {M.}~\bibnamefont {Niethammer}}, \bibinfo {author} {\bibfnamefont {D.}~\bibnamefont {Dasari}}, \bibinfo {author} {\bibfnamefont {V.}~\bibnamefont {Vorobyov}}, \emph {et~al.},\ }\bibfield  {title} {\bibinfo {title} {Spin-controlled generation of indistinguishable and distinguishable photons from silicon vacancy centres in silicon carbide},\ }\href@noop {} {\bibfield  {journal} {\bibinfo  {journal} {Nature communications}\ }\textbf {\bibinfo {volume} {11}},\ \bibinfo {pages} {2516} (\bibinfo {year}
  {2020}{\natexlab{a}})}\BibitemShut {NoStop}%
\bibitem [{\citenamefont {Fang}\ \emph {et~al.}(2024)\citenamefont {Fang}, \citenamefont {Lai}, \citenamefont {Li}, \citenamefont {Su}, \citenamefont {Lu}, \citenamefont {Yang}, \citenamefont {Liu}, \citenamefont {Qiao}, \citenamefont {Li}, \citenamefont {He} \emph {et~al.}}]{fang2024experimental}%
  \BibitemOpen
  \bibfield  {author} {\bibinfo {author} {\bibfnamefont {R.-Z.}\ \bibnamefont {Fang}}, \bibinfo {author} {\bibfnamefont {X.-Y.}\ \bibnamefont {Lai}}, \bibinfo {author} {\bibfnamefont {T.}~\bibnamefont {Li}}, \bibinfo {author} {\bibfnamefont {R.-Z.}\ \bibnamefont {Su}}, \bibinfo {author} {\bibfnamefont {B.-W.}\ \bibnamefont {Lu}}, \bibinfo {author} {\bibfnamefont {C.-W.}\ \bibnamefont {Yang}}, \bibinfo {author} {\bibfnamefont {R.-Z.}\ \bibnamefont {Liu}}, \bibinfo {author} {\bibfnamefont {Y.-K.}\ \bibnamefont {Qiao}}, \bibinfo {author} {\bibfnamefont {C.}~\bibnamefont {Li}}, \bibinfo {author} {\bibfnamefont {Z.-G.}\ \bibnamefont {He}}, \emph {et~al.},\ }\bibfield  {title} {\bibinfo {title} {Experimental generation of spin-photon entanglement in silicon carbide},\ }\href@noop {} {\bibfield  {journal} {\bibinfo  {journal} {Physical Review Letters}\ }\textbf {\bibinfo {volume} {132}},\ \bibinfo {pages} {160801} (\bibinfo {year} {2024})}\BibitemShut {NoStop}%
\bibitem [{\citenamefont {Nguyen}\ \emph {et~al.}(2019)\citenamefont {Nguyen}, \citenamefont {Sukachev}, \citenamefont {Bhaskar}, \citenamefont {Machielse}, \citenamefont {Levonian}, \citenamefont {Knall}, \citenamefont {Stroganov}, \citenamefont {Chia}, \citenamefont {Burek}, \citenamefont {Riedinger} \emph {et~al.}}]{nguyen2019integrated}%
  \BibitemOpen
  \bibfield  {author} {\bibinfo {author} {\bibfnamefont {C.}~\bibnamefont {Nguyen}}, \bibinfo {author} {\bibfnamefont {D.}~\bibnamefont {Sukachev}}, \bibinfo {author} {\bibfnamefont {M.}~\bibnamefont {Bhaskar}}, \bibinfo {author} {\bibfnamefont {B.}~\bibnamefont {Machielse}}, \bibinfo {author} {\bibfnamefont {D.}~\bibnamefont {Levonian}}, \bibinfo {author} {\bibfnamefont {E.}~\bibnamefont {Knall}}, \bibinfo {author} {\bibfnamefont {P.}~\bibnamefont {Stroganov}}, \bibinfo {author} {\bibfnamefont {C.}~\bibnamefont {Chia}}, \bibinfo {author} {\bibfnamefont {M.}~\bibnamefont {Burek}}, \bibinfo {author} {\bibfnamefont {R.}~\bibnamefont {Riedinger}}, \emph {et~al.},\ }\bibfield  {title} {\bibinfo {title} {An integrated nanophotonic quantum register based on silicon-vacancy spins in diamond},\ }\href@noop {} {\bibfield  {journal} {\bibinfo  {journal} {Physical Review B}\ }\textbf {\bibinfo {volume} {100}},\ \bibinfo {pages} {165428} (\bibinfo {year} {2019})}\BibitemShut {NoStop}%
\bibitem [{\citenamefont {Knaut}\ \emph {et~al.}(2024)\citenamefont {Knaut}, \citenamefont {Suleymanzade}, \citenamefont {Wei}, \citenamefont {Assumpcao}, \citenamefont {Stas}, \citenamefont {Huan}, \citenamefont {Machielse}, \citenamefont {Knall}, \citenamefont {Sutula}, \citenamefont {Baranes}, \citenamefont {Sinclair}, \citenamefont {De-Eknamkul}, \citenamefont {Levonian}, \citenamefont {Bhaskar}, \citenamefont {Park}, \citenamefont {Lon{\v c}ar},\ and\ \citenamefont {Lukin}}]{knaut_entanglement_2024}%
  \BibitemOpen
  \bibfield  {author} {\bibinfo {author} {\bibfnamefont {C.~M.}\ \bibnamefont {Knaut}}, \bibinfo {author} {\bibfnamefont {A.}~\bibnamefont {Suleymanzade}}, \bibinfo {author} {\bibfnamefont {Y.-C.}\ \bibnamefont {Wei}}, \bibinfo {author} {\bibfnamefont {D.~R.}\ \bibnamefont {Assumpcao}}, \bibinfo {author} {\bibfnamefont {P.-J.}\ \bibnamefont {Stas}}, \bibinfo {author} {\bibfnamefont {Y.~Q.}\ \bibnamefont {Huan}}, \bibinfo {author} {\bibfnamefont {B.}~\bibnamefont {Machielse}}, \bibinfo {author} {\bibfnamefont {E.~N.}\ \bibnamefont {Knall}}, \bibinfo {author} {\bibfnamefont {M.}~\bibnamefont {Sutula}}, \bibinfo {author} {\bibfnamefont {G.}~\bibnamefont {Baranes}}, \bibinfo {author} {\bibfnamefont {N.}~\bibnamefont {Sinclair}}, \bibinfo {author} {\bibfnamefont {C.}~\bibnamefont {De-Eknamkul}}, \bibinfo {author} {\bibfnamefont {D.~S.}\ \bibnamefont {Levonian}}, \bibinfo {author} {\bibfnamefont {M.~K.}\ \bibnamefont {Bhaskar}}, \bibinfo {author} {\bibfnamefont {H.}~\bibnamefont {Park}}, \bibinfo {author}
  {\bibfnamefont {M.}~\bibnamefont {Lon{\v c}ar}},\ and\ \bibinfo {author} {\bibfnamefont {M.~D.}\ \bibnamefont {Lukin}},\ }\bibfield  {title} {\bibinfo {title} {Entanglement of nanophotonic quantum memory nodes in a telecom network},\ }\href {https://doi.org/10.1038/s41586-024-07252-z} {\bibfield  {journal} {\bibinfo  {journal} {Nature}\ }\textbf {\bibinfo {volume} {629}},\ \bibinfo {pages} {573} (\bibinfo {year} {2024})},\ \bibinfo {note} {publisher: Nature Publishing Group}\BibitemShut {NoStop}%
\bibitem [{\citenamefont {Sukachev}\ \emph {et~al.}(2017{\natexlab{a}})\citenamefont {Sukachev}, \citenamefont {Sipahigil}, \citenamefont {Nguyen}, \citenamefont {Bhaskar}, \citenamefont {Evans}, \citenamefont {Jelezko},\ and\ \citenamefont {Lukin}}]{sukachev2017silicon}%
  \BibitemOpen
  \bibfield  {author} {\bibinfo {author} {\bibfnamefont {D.~D.}\ \bibnamefont {Sukachev}}, \bibinfo {author} {\bibfnamefont {A.}~\bibnamefont {Sipahigil}}, \bibinfo {author} {\bibfnamefont {C.~T.}\ \bibnamefont {Nguyen}}, \bibinfo {author} {\bibfnamefont {M.~K.}\ \bibnamefont {Bhaskar}}, \bibinfo {author} {\bibfnamefont {R.~E.}\ \bibnamefont {Evans}}, \bibinfo {author} {\bibfnamefont {F.}~\bibnamefont {Jelezko}},\ and\ \bibinfo {author} {\bibfnamefont {M.~D.}\ \bibnamefont {Lukin}},\ }\bibfield  {title} {\bibinfo {title} {Silicon-vacancy spin qubit in diamond: a quantum memory exceeding 10 ms with single-shot state readout},\ }\href@noop {} {\bibfield  {journal} {\bibinfo  {journal} {Physical review letters}\ }\textbf {\bibinfo {volume} {119}},\ \bibinfo {pages} {223602} (\bibinfo {year} {2017}{\natexlab{a}})}\BibitemShut {NoStop}%
\bibitem [{\citenamefont {Stas}\ \emph {et~al.}(2022)\citenamefont {Stas}, \citenamefont {Huan}, \citenamefont {Machielse}, \citenamefont {Knall}, \citenamefont {Suleymanzade}, \citenamefont {Pingault}, \citenamefont {Sutula}, \citenamefont {Ding}, \citenamefont {Knaut}, \citenamefont {Assumpcao}, \citenamefont {Wei}, \citenamefont {Bhaskar}, \citenamefont {Riedinger}, \citenamefont {Sukachev}, \citenamefont {Park}, \citenamefont {Lon{\v c}ar}, \citenamefont {Levonian},\ and\ \citenamefont {Lukin}}]{stas_robust_2022}%
  \BibitemOpen
  \bibfield  {author} {\bibinfo {author} {\bibfnamefont {P.-J.}\ \bibnamefont {Stas}}, \bibinfo {author} {\bibfnamefont {Y.~Q.}\ \bibnamefont {Huan}}, \bibinfo {author} {\bibfnamefont {B.}~\bibnamefont {Machielse}}, \bibinfo {author} {\bibfnamefont {E.~N.}\ \bibnamefont {Knall}}, \bibinfo {author} {\bibfnamefont {A.}~\bibnamefont {Suleymanzade}}, \bibinfo {author} {\bibfnamefont {B.}~\bibnamefont {Pingault}}, \bibinfo {author} {\bibfnamefont {M.}~\bibnamefont {Sutula}}, \bibinfo {author} {\bibfnamefont {S.~W.}\ \bibnamefont {Ding}}, \bibinfo {author} {\bibfnamefont {C.~M.}\ \bibnamefont {Knaut}}, \bibinfo {author} {\bibfnamefont {D.~R.}\ \bibnamefont {Assumpcao}}, \bibinfo {author} {\bibfnamefont {Y.-C.}\ \bibnamefont {Wei}}, \bibinfo {author} {\bibfnamefont {M.~K.}\ \bibnamefont {Bhaskar}}, \bibinfo {author} {\bibfnamefont {R.}~\bibnamefont {Riedinger}}, \bibinfo {author} {\bibfnamefont {D.~D.}\ \bibnamefont {Sukachev}}, \bibinfo {author} {\bibfnamefont {H.}~\bibnamefont {Park}}, \bibinfo {author}
  {\bibfnamefont {M.}~\bibnamefont {Lon{\v c}ar}}, \bibinfo {author} {\bibfnamefont {D.~S.}\ \bibnamefont {Levonian}},\ and\ \bibinfo {author} {\bibfnamefont {M.~D.}\ \bibnamefont {Lukin}},\ }\bibfield  {title} {\bibinfo {title} {Robust multi-qubit quantum network node with integrated error detection},\ }\href {https://doi.org/10.1126/science.add9771} {\bibfield  {journal} {\bibinfo  {journal} {Science}\ }\textbf {\bibinfo {volume} {378}},\ \bibinfo {pages} {557} (\bibinfo {year} {2022})},\ \bibinfo {note} {publisher: American Association for the Advancement of Science}\BibitemShut {NoStop}%
\bibitem [{\citenamefont {Hepp}\ \emph {et~al.}(2014)\citenamefont {Hepp}, \citenamefont {M{\"u}ller}, \citenamefont {Waselowski}, \citenamefont {Becker}, \citenamefont {Pingault}, \citenamefont {Sternschulte}, \citenamefont {Steinm{\"u}ller-Nethl}, \citenamefont {Gali}, \citenamefont {Maze}, \citenamefont {Atat{\"u}re},\ and\ \citenamefont {Becher}}]{hepp_electronic_2014}%
  \BibitemOpen
  \bibfield  {author} {\bibinfo {author} {\bibfnamefont {C.}~\bibnamefont {Hepp}}, \bibinfo {author} {\bibfnamefont {T.}~\bibnamefont {M{\"u}ller}}, \bibinfo {author} {\bibfnamefont {V.}~\bibnamefont {Waselowski}}, \bibinfo {author} {\bibfnamefont {J.~N.}\ \bibnamefont {Becker}}, \bibinfo {author} {\bibfnamefont {B.}~\bibnamefont {Pingault}}, \bibinfo {author} {\bibfnamefont {H.}~\bibnamefont {Sternschulte}}, \bibinfo {author} {\bibfnamefont {D.}~\bibnamefont {Steinm{\"u}ller-Nethl}}, \bibinfo {author} {\bibfnamefont {A.}~\bibnamefont {Gali}}, \bibinfo {author} {\bibfnamefont {J.~R.}\ \bibnamefont {Maze}}, \bibinfo {author} {\bibfnamefont {M.}~\bibnamefont {Atat{\"u}re}},\ and\ \bibinfo {author} {\bibfnamefont {C.}~\bibnamefont {Becher}},\ }\bibfield  {title} {\bibinfo {title} {Electronic {Structure} of the {Silicon} {Vacancy} {Color} {Center} in {Diamond}},\ }\href {https://doi.org/10.1103/PhysRevLett.112.036405} {\bibfield  {journal} {\bibinfo  {journal} {Physical Review Letters}\ }\textbf {\bibinfo
  {volume} {112}},\ \bibinfo {pages} {036405} (\bibinfo {year} {2014})},\ \bibinfo {note} {publisher: American Physical Society}\BibitemShut {NoStop}%
\bibitem [{\citenamefont {Thiering}\ and\ \citenamefont {Gali}(2018)}]{thiering_ab_2018}%
  \BibitemOpen
  \bibfield  {author} {\bibinfo {author} {\bibfnamefont {G.}~\bibnamefont {Thiering}}\ and\ \bibinfo {author} {\bibfnamefont {A.}~\bibnamefont {Gali}},\ }\bibfield  {title} {\bibinfo {title} {\textit{{Ab} {Initio}} {Magneto}-{Optical} {Spectrum} of {Group}-{IV} {Vacancy} {Color} {Centers} in {Diamond}},\ }\href {https://doi.org/10.1103/PhysRevX.8.021063} {\bibfield  {journal} {\bibinfo  {journal} {Physical Review X}\ }\textbf {\bibinfo {volume} {8}},\ \bibinfo {pages} {021063} (\bibinfo {year} {2018})}\BibitemShut {NoStop}%
\bibitem [{\citenamefont {Sipahigil}\ \emph {et~al.}(2016)\citenamefont {Sipahigil}, \citenamefont {Evans}, \citenamefont {Sukachev}, \citenamefont {Burek}, \citenamefont {Borregaard}, \citenamefont {Bhaskar}, \citenamefont {Nguyen}, \citenamefont {Pacheco}, \citenamefont {Atikian}, \citenamefont {Meuwly}, \citenamefont {Camacho}, \citenamefont {Jelezko}, \citenamefont {Bielejec}, \citenamefont {Park}, \citenamefont {Lon{\v c}ar},\ and\ \citenamefont {Lukin}}]{sipahigil_integrated_2016}%
  \BibitemOpen
  \bibfield  {author} {\bibinfo {author} {\bibfnamefont {A.}~\bibnamefont {Sipahigil}}, \bibinfo {author} {\bibfnamefont {R.~E.}\ \bibnamefont {Evans}}, \bibinfo {author} {\bibfnamefont {D.~D.}\ \bibnamefont {Sukachev}}, \bibinfo {author} {\bibfnamefont {M.~J.}\ \bibnamefont {Burek}}, \bibinfo {author} {\bibfnamefont {J.}~\bibnamefont {Borregaard}}, \bibinfo {author} {\bibfnamefont {M.~K.}\ \bibnamefont {Bhaskar}}, \bibinfo {author} {\bibfnamefont {C.~T.}\ \bibnamefont {Nguyen}}, \bibinfo {author} {\bibfnamefont {J.~L.}\ \bibnamefont {Pacheco}}, \bibinfo {author} {\bibfnamefont {H.~A.}\ \bibnamefont {Atikian}}, \bibinfo {author} {\bibfnamefont {C.}~\bibnamefont {Meuwly}}, \bibinfo {author} {\bibfnamefont {R.~M.}\ \bibnamefont {Camacho}}, \bibinfo {author} {\bibfnamefont {F.}~\bibnamefont {Jelezko}}, \bibinfo {author} {\bibfnamefont {E.}~\bibnamefont {Bielejec}}, \bibinfo {author} {\bibfnamefont {H.}~\bibnamefont {Park}}, \bibinfo {author} {\bibfnamefont {M.}~\bibnamefont {Lon{\v c}ar}},\ and\ \bibinfo
  {author} {\bibfnamefont {M.~D.}\ \bibnamefont {Lukin}},\ }\bibfield  {title} {\bibinfo {title} {An integrated diamond nanophotonics platform for quantum-optical networks},\ }\href {https://doi.org/10.1126/science.aah6875} {\bibfield  {journal} {\bibinfo  {journal} {Science}\ }\textbf {\bibinfo {volume} {354}},\ \bibinfo {pages} {847} (\bibinfo {year} {2016})}\BibitemShut {NoStop}%
\bibitem [{\citenamefont {Evans}\ \emph {et~al.}(2016)\citenamefont {Evans}, \citenamefont {Sipahigil}, \citenamefont {Sukachev}, \citenamefont {Zibrov},\ and\ \citenamefont {Lukin}}]{evans2016narrow}%
  \BibitemOpen
  \bibfield  {author} {\bibinfo {author} {\bibfnamefont {R.~E.}\ \bibnamefont {Evans}}, \bibinfo {author} {\bibfnamefont {A.}~\bibnamefont {Sipahigil}}, \bibinfo {author} {\bibfnamefont {D.~D.}\ \bibnamefont {Sukachev}}, \bibinfo {author} {\bibfnamefont {A.~S.}\ \bibnamefont {Zibrov}},\ and\ \bibinfo {author} {\bibfnamefont {M.~D.}\ \bibnamefont {Lukin}},\ }\bibfield  {title} {\bibinfo {title} {Narrow-linewidth homogeneous optical emitters in diamond nanostructures via silicon ion implantation},\ }\href@noop {} {\bibfield  {journal} {\bibinfo  {journal} {Physical Review Applied}\ }\textbf {\bibinfo {volume} {5}},\ \bibinfo {pages} {044010} (\bibinfo {year} {2016})}\BibitemShut {NoStop}%
\bibitem [{\citenamefont {Schr{\"o}der}\ \emph {et~al.}(2017)\citenamefont {Schr{\"o}der}, \citenamefont {Trusheim}, \citenamefont {Walsh}, \citenamefont {Li}, \citenamefont {Zheng}, \citenamefont {Schukraft}, \citenamefont {Sipahigil}, \citenamefont {Evans}, \citenamefont {Sukachev}, \citenamefont {Nguyen} \emph {et~al.}}]{schroder2017scalable}%
  \BibitemOpen
  \bibfield  {author} {\bibinfo {author} {\bibfnamefont {T.}~\bibnamefont {Schr{\"o}der}}, \bibinfo {author} {\bibfnamefont {M.~E.}\ \bibnamefont {Trusheim}}, \bibinfo {author} {\bibfnamefont {M.}~\bibnamefont {Walsh}}, \bibinfo {author} {\bibfnamefont {L.}~\bibnamefont {Li}}, \bibinfo {author} {\bibfnamefont {J.}~\bibnamefont {Zheng}}, \bibinfo {author} {\bibfnamefont {M.}~\bibnamefont {Schukraft}}, \bibinfo {author} {\bibfnamefont {A.}~\bibnamefont {Sipahigil}}, \bibinfo {author} {\bibfnamefont {R.~E.}\ \bibnamefont {Evans}}, \bibinfo {author} {\bibfnamefont {D.~D.}\ \bibnamefont {Sukachev}}, \bibinfo {author} {\bibfnamefont {C.~T.}\ \bibnamefont {Nguyen}}, \emph {et~al.},\ }\bibfield  {title} {\bibinfo {title} {Scalable focused ion beam creation of nearly lifetime-limited single quantum emitters in diamond nanostructures},\ }\href@noop {} {\bibfield  {journal} {\bibinfo  {journal} {Nature communications}\ }\textbf {\bibinfo {volume} {8}},\ \bibinfo {pages} {15376} (\bibinfo {year} {2017})}\BibitemShut
  {NoStop}%
\bibitem [{\citenamefont {Sipahigil}\ \emph {et~al.}(2014)\citenamefont {Sipahigil}, \citenamefont {Jahnke}, \citenamefont {Rogers}, \citenamefont {Teraji}, \citenamefont {Isoya}, \citenamefont {Zibrov}, \citenamefont {Jelezko},\ and\ \citenamefont {Lukin}}]{sipahigil_indistinguishable_2014}%
  \BibitemOpen
  \bibfield  {author} {\bibinfo {author} {\bibfnamefont {A.}~\bibnamefont {Sipahigil}}, \bibinfo {author} {\bibfnamefont {K.}~\bibnamefont {Jahnke}}, \bibinfo {author} {\bibfnamefont {L.}~\bibnamefont {Rogers}}, \bibinfo {author} {\bibfnamefont {T.}~\bibnamefont {Teraji}}, \bibinfo {author} {\bibfnamefont {J.}~\bibnamefont {Isoya}}, \bibinfo {author} {\bibfnamefont {A.}~\bibnamefont {Zibrov}}, \bibinfo {author} {\bibfnamefont {F.}~\bibnamefont {Jelezko}},\ and\ \bibinfo {author} {\bibfnamefont {M.}~\bibnamefont {Lukin}},\ }\bibfield  {title} {\bibinfo {title} {Indistinguishable {Photons} from {Separated} {Silicon}-{Vacancy} {Centers} in {Diamond}},\ }\href {https://doi.org/10.1103/PhysRevLett.113.113602} {\bibfield  {journal} {\bibinfo  {journal} {Physical Review Letters}\ }\textbf {\bibinfo {volume} {113}},\ \bibinfo {pages} {113602} (\bibinfo {year} {2014})}\BibitemShut {NoStop}%
\bibitem [{\citenamefont {Bhaskar}\ \emph {et~al.}(2020)\citenamefont {Bhaskar}, \citenamefont {Riedinger}, \citenamefont {Machielse}, \citenamefont {Levonian}, \citenamefont {Nguyen}, \citenamefont {Knall}, \citenamefont {Park}, \citenamefont {Englund}, \citenamefont {Lon{\v c}ar}, \citenamefont {Sukachev},\ and\ \citenamefont {Lukin}}]{bhaskar_experimental_2020}%
  \BibitemOpen
  \bibfield  {author} {\bibinfo {author} {\bibfnamefont {M.~K.}\ \bibnamefont {Bhaskar}}, \bibinfo {author} {\bibfnamefont {R.}~\bibnamefont {Riedinger}}, \bibinfo {author} {\bibfnamefont {B.}~\bibnamefont {Machielse}}, \bibinfo {author} {\bibfnamefont {D.~S.}\ \bibnamefont {Levonian}}, \bibinfo {author} {\bibfnamefont {C.~T.}\ \bibnamefont {Nguyen}}, \bibinfo {author} {\bibfnamefont {E.~N.}\ \bibnamefont {Knall}}, \bibinfo {author} {\bibfnamefont {H.}~\bibnamefont {Park}}, \bibinfo {author} {\bibfnamefont {D.}~\bibnamefont {Englund}}, \bibinfo {author} {\bibfnamefont {M.}~\bibnamefont {Lon{\v c}ar}}, \bibinfo {author} {\bibfnamefont {D.~D.}\ \bibnamefont {Sukachev}},\ and\ \bibinfo {author} {\bibfnamefont {M.~D.}\ \bibnamefont {Lukin}},\ }\bibfield  {title} {\bibinfo {title} {Experimental demonstration of memory-enhanced quantum communication},\ }\href {https://doi.org/10.1038/s41586-020-2103-5} {\bibfield  {journal} {\bibinfo  {journal} {Nature}\ }\textbf {\bibinfo {volume} {580}},\ \bibinfo {pages} {60}
  (\bibinfo {year} {2020})}\BibitemShut {NoStop}%
\bibitem [{\citenamefont {Jahnke}\ \emph {et~al.}(2015)\citenamefont {Jahnke}, \citenamefont {Sipahigil}, \citenamefont {Binder}, \citenamefont {Doherty}, \citenamefont {Metsch}, \citenamefont {Rogers}, \citenamefont {Manson}, \citenamefont {Lukin},\ and\ \citenamefont {Jelezko}}]{jahnke_electronphonon_2015}%
  \BibitemOpen
  \bibfield  {author} {\bibinfo {author} {\bibfnamefont {K.~D.}\ \bibnamefont {Jahnke}}, \bibinfo {author} {\bibfnamefont {A.}~\bibnamefont {Sipahigil}}, \bibinfo {author} {\bibfnamefont {J.~M.}\ \bibnamefont {Binder}}, \bibinfo {author} {\bibfnamefont {M.~W.}\ \bibnamefont {Doherty}}, \bibinfo {author} {\bibfnamefont {M.}~\bibnamefont {Metsch}}, \bibinfo {author} {\bibfnamefont {L.~J.}\ \bibnamefont {Rogers}}, \bibinfo {author} {\bibfnamefont {N.~B.}\ \bibnamefont {Manson}}, \bibinfo {author} {\bibfnamefont {M.~D.}\ \bibnamefont {Lukin}},\ and\ \bibinfo {author} {\bibfnamefont {F.}~\bibnamefont {Jelezko}},\ }\bibfield  {title} {\bibinfo {title} {Electron--phonon processes of the silicon-vacancy centre in diamond},\ }\href {https://doi.org/10.1088/1367-2630/17/4/043011} {\bibfield  {journal} {\bibinfo  {journal} {New Journal of Physics}\ }\textbf {\bibinfo {volume} {17}},\ \bibinfo {pages} {043011} (\bibinfo {year} {2015})}\BibitemShut {NoStop}%
\bibitem [{\citenamefont {Sukachev}\ \emph {et~al.}(2017{\natexlab{b}})\citenamefont {Sukachev}, \citenamefont {Sipahigil}, \citenamefont {Nguyen}, \citenamefont {Bhaskar}, \citenamefont {Evans}, \citenamefont {Jelezko},\ and\ \citenamefont {Lukin}}]{sukachev_silicon-vacancy_2017}%
  \BibitemOpen
  \bibfield  {author} {\bibinfo {author} {\bibfnamefont {D.}~\bibnamefont {Sukachev}}, \bibinfo {author} {\bibfnamefont {A.}~\bibnamefont {Sipahigil}}, \bibinfo {author} {\bibfnamefont {C.}~\bibnamefont {Nguyen}}, \bibinfo {author} {\bibfnamefont {M.}~\bibnamefont {Bhaskar}}, \bibinfo {author} {\bibfnamefont {R.}~\bibnamefont {Evans}}, \bibinfo {author} {\bibfnamefont {F.}~\bibnamefont {Jelezko}},\ and\ \bibinfo {author} {\bibfnamefont {M.}~\bibnamefont {Lukin}},\ }\bibfield  {title} {\bibinfo {title} {Silicon-{Vacancy} {Spin} {Qubit} in {Diamond}: {A} {Quantum} {Memory} {Exceeding} 10 ms with {Single}-{Shot} {State} {Readout}},\ }\href {https://doi.org/10.1103/PhysRevLett.119.223602} {\bibfield  {journal} {\bibinfo  {journal} {Physical Review Letters}\ }\textbf {\bibinfo {volume} {119}},\ \bibinfo {pages} {223602} (\bibinfo {year} {2017}{\natexlab{b}})},\ \bibinfo {note} {publisher: American Physical Society}\BibitemShut {NoStop}%
\bibitem [{\citenamefont {Siyushev}\ \emph {et~al.}(2017)\citenamefont {Siyushev}, \citenamefont {Metsch}, \citenamefont {Ijaz}, \citenamefont {Binder}, \citenamefont {Bhaskar}, \citenamefont {Sukachev}, \citenamefont {Sipahigil}, \citenamefont {Evans}, \citenamefont {Nguyen}, \citenamefont {Lukin}, \citenamefont {Hemmer}, \citenamefont {Palyanov}, \citenamefont {Kupriyanov}, \citenamefont {Borzdov}, \citenamefont {Rogers},\ and\ \citenamefont {Jelezko}}]{siyushev_optical_2017}%
  \BibitemOpen
  \bibfield  {author} {\bibinfo {author} {\bibfnamefont {P.}~\bibnamefont {Siyushev}}, \bibinfo {author} {\bibfnamefont {M.~H.}\ \bibnamefont {Metsch}}, \bibinfo {author} {\bibfnamefont {A.}~\bibnamefont {Ijaz}}, \bibinfo {author} {\bibfnamefont {J.~M.}\ \bibnamefont {Binder}}, \bibinfo {author} {\bibfnamefont {M.~K.}\ \bibnamefont {Bhaskar}}, \bibinfo {author} {\bibfnamefont {D.~D.}\ \bibnamefont {Sukachev}}, \bibinfo {author} {\bibfnamefont {A.}~\bibnamefont {Sipahigil}}, \bibinfo {author} {\bibfnamefont {R.~E.}\ \bibnamefont {Evans}}, \bibinfo {author} {\bibfnamefont {C.~T.}\ \bibnamefont {Nguyen}}, \bibinfo {author} {\bibfnamefont {M.~D.}\ \bibnamefont {Lukin}}, \bibinfo {author} {\bibfnamefont {P.~R.}\ \bibnamefont {Hemmer}}, \bibinfo {author} {\bibfnamefont {Y.~N.}\ \bibnamefont {Palyanov}}, \bibinfo {author} {\bibfnamefont {I.~N.}\ \bibnamefont {Kupriyanov}}, \bibinfo {author} {\bibfnamefont {Y.~M.}\ \bibnamefont {Borzdov}}, \bibinfo {author} {\bibfnamefont {L.~J.}\ \bibnamefont {Rogers}},\ and\
  \bibinfo {author} {\bibfnamefont {F.}~\bibnamefont {Jelezko}},\ }\bibfield  {title} {\bibinfo {title} {Optical and microwave control of germanium-vacancy center spins in diamond},\ }\href {https://doi.org/10.1103/PhysRevB.96.081201} {\bibfield  {journal} {\bibinfo  {journal} {Physical Review B}\ }\textbf {\bibinfo {volume} {96}},\ \bibinfo {pages} {081201} (\bibinfo {year} {2017})}\BibitemShut {NoStop}%
\bibitem [{\citenamefont {Iwasaki}\ \emph {et~al.}(2017)\citenamefont {Iwasaki}, \citenamefont {Miyamoto}, \citenamefont {Taniguchi}, \citenamefont {Siyushev}, \citenamefont {Metsch}, \citenamefont {Jelezko},\ and\ \citenamefont {Hatano}}]{iwasaki_tin-vacancy_2017}%
  \BibitemOpen
  \bibfield  {author} {\bibinfo {author} {\bibfnamefont {T.}~\bibnamefont {Iwasaki}}, \bibinfo {author} {\bibfnamefont {Y.}~\bibnamefont {Miyamoto}}, \bibinfo {author} {\bibfnamefont {T.}~\bibnamefont {Taniguchi}}, \bibinfo {author} {\bibfnamefont {P.}~\bibnamefont {Siyushev}}, \bibinfo {author} {\bibfnamefont {M.~H.}\ \bibnamefont {Metsch}}, \bibinfo {author} {\bibfnamefont {F.}~\bibnamefont {Jelezko}},\ and\ \bibinfo {author} {\bibfnamefont {M.}~\bibnamefont {Hatano}},\ }\bibfield  {title} {\bibinfo {title} {Tin-{Vacancy} {Quantum} {Emitters} in {Diamond}},\ }\href {https://doi.org/10.1103/PhysRevLett.119.253601} {\bibfield  {journal} {\bibinfo  {journal} {Physical Review Letters}\ }\textbf {\bibinfo {volume} {119}},\ \bibinfo {pages} {253601} (\bibinfo {year} {2017})}\BibitemShut {NoStop}%
\bibitem [{\citenamefont {Trusheim}\ \emph {et~al.}(2020)\citenamefont {Trusheim}, \citenamefont {Pingault}, \citenamefont {Wan}, \citenamefont {G{\"u}ndo{\u g}an}, \citenamefont {De~Santis}, \citenamefont {Debroux}, \citenamefont {Gangloff}, \citenamefont {Purser}, \citenamefont {Chen}, \citenamefont {Walsh}, \citenamefont {Rose}, \citenamefont {Becker}, \citenamefont {Lienhard}, \citenamefont {Bersin}, \citenamefont {Paradeisanos}, \citenamefont {Wang}, \citenamefont {Lyzwa}, \citenamefont {Montblanch}, \citenamefont {Malladi}, \citenamefont {Bakhru}, \citenamefont {Ferrari}, \citenamefont {Walmsley}, \citenamefont {Atat{\"u}re},\ and\ \citenamefont {Englund}}]{trusheim_transform-limited_2020}%
  \BibitemOpen
  \bibfield  {author} {\bibinfo {author} {\bibfnamefont {M.~E.}\ \bibnamefont {Trusheim}}, \bibinfo {author} {\bibfnamefont {B.}~\bibnamefont {Pingault}}, \bibinfo {author} {\bibfnamefont {N.~H.}\ \bibnamefont {Wan}}, \bibinfo {author} {\bibfnamefont {M.}~\bibnamefont {G{\"u}ndo{\u g}an}}, \bibinfo {author} {\bibfnamefont {L.}~\bibnamefont {De~Santis}}, \bibinfo {author} {\bibfnamefont {R.}~\bibnamefont {Debroux}}, \bibinfo {author} {\bibfnamefont {D.}~\bibnamefont {Gangloff}}, \bibinfo {author} {\bibfnamefont {C.}~\bibnamefont {Purser}}, \bibinfo {author} {\bibfnamefont {K.~C.}\ \bibnamefont {Chen}}, \bibinfo {author} {\bibfnamefont {M.}~\bibnamefont {Walsh}}, \bibinfo {author} {\bibfnamefont {J.~J.}\ \bibnamefont {Rose}}, \bibinfo {author} {\bibfnamefont {J.~N.}\ \bibnamefont {Becker}}, \bibinfo {author} {\bibfnamefont {B.}~\bibnamefont {Lienhard}}, \bibinfo {author} {\bibfnamefont {E.}~\bibnamefont {Bersin}}, \bibinfo {author} {\bibfnamefont {I.}~\bibnamefont {Paradeisanos}}, \bibinfo {author}
  {\bibfnamefont {G.}~\bibnamefont {Wang}}, \bibinfo {author} {\bibfnamefont {D.}~\bibnamefont {Lyzwa}}, \bibinfo {author} {\bibfnamefont {A.~R.-P.}\ \bibnamefont {Montblanch}}, \bibinfo {author} {\bibfnamefont {G.}~\bibnamefont {Malladi}}, \bibinfo {author} {\bibfnamefont {H.}~\bibnamefont {Bakhru}}, \bibinfo {author} {\bibfnamefont {A.~C.}\ \bibnamefont {Ferrari}}, \bibinfo {author} {\bibfnamefont {I.~A.}\ \bibnamefont {Walmsley}}, \bibinfo {author} {\bibfnamefont {M.}~\bibnamefont {Atat{\"u}re}},\ and\ \bibinfo {author} {\bibfnamefont {D.}~\bibnamefont {Englund}},\ }\bibfield  {title} {\bibinfo {title} {Transform-{Limited} {Photons} {From} a {Coherent} {Tin}-{Vacancy} {Spin} in {Diamond}},\ }\href {https://doi.org/10.1103/PhysRevLett.124.023602} {\bibfield  {journal} {\bibinfo  {journal} {Physical Review Letters}\ }\textbf {\bibinfo {volume} {124}},\ \bibinfo {pages} {023602} (\bibinfo {year} {2020})}\BibitemShut {NoStop}%
\bibitem [{\citenamefont {Senkalla}\ \emph {et~al.}(2024)\citenamefont {Senkalla}, \citenamefont {Genov}, \citenamefont {Metsch}, \citenamefont {Siyushev},\ and\ \citenamefont {Jelezko}}]{senkalla_germanium_2024}%
  \BibitemOpen
  \bibfield  {author} {\bibinfo {author} {\bibfnamefont {K.}~\bibnamefont {Senkalla}}, \bibinfo {author} {\bibfnamefont {G.}~\bibnamefont {Genov}}, \bibinfo {author} {\bibfnamefont {M.~H.}\ \bibnamefont {Metsch}}, \bibinfo {author} {\bibfnamefont {P.}~\bibnamefont {Siyushev}},\ and\ \bibinfo {author} {\bibfnamefont {F.}~\bibnamefont {Jelezko}},\ }\bibfield  {title} {\bibinfo {title} {Germanium {Vacancy} in {Diamond} {Quantum} {Memory} {Exceeding} 20 ms},\ }\href {https://doi.org/10.1103/PhysRevLett.132.026901} {\bibfield  {journal} {\bibinfo  {journal} {Physical Review Letters}\ }\textbf {\bibinfo {volume} {132}},\ \bibinfo {pages} {026901} (\bibinfo {year} {2024})}\BibitemShut {NoStop}%
\bibitem [{\citenamefont {Zahedian}\ \emph {et~al.}(2024)\citenamefont {Zahedian}, \citenamefont {Vorobyov},\ and\ \citenamefont {Wrachtrup}}]{zahedian_blueprint_2024}%
  \BibitemOpen
  \bibfield  {author} {\bibinfo {author} {\bibfnamefont {M.}~\bibnamefont {Zahedian}}, \bibinfo {author} {\bibfnamefont {V.}~\bibnamefont {Vorobyov}},\ and\ \bibinfo {author} {\bibfnamefont {J.}~\bibnamefont {Wrachtrup}},\ }\bibfield  {title} {\bibinfo {title} {Blueprint for efficient nuclear spin characterization with color centers},\ }\href {https://doi.org/10.1103/PhysRevB.109.214111} {\bibfield  {journal} {\bibinfo  {journal} {Physical Review B}\ }\textbf {\bibinfo {volume} {109}},\ \bibinfo {pages} {214111} (\bibinfo {year} {2024})}\BibitemShut {NoStop}%
\bibitem [{\citenamefont {Guo}\ \emph {et~al.}(2023)\citenamefont {Guo}, \citenamefont {Stramma}, \citenamefont {Li}, \citenamefont {Roth}, \citenamefont {Huang}, \citenamefont {Jin}, \citenamefont {Parker}, \citenamefont {Arjona~Mart{\'\i}nez}, \citenamefont {Shofer}, \citenamefont {Michaels}, \citenamefont {Purser}, \citenamefont {Appel}, \citenamefont {Alexeev}, \citenamefont {Liu}, \citenamefont {Ferrari}, \citenamefont {Awschalom}, \citenamefont {Delegan}, \citenamefont {Pingault}, \citenamefont {Galli}, \citenamefont {Heremans}, \citenamefont {Atat{\"u}re},\ and\ \citenamefont {High}}]{guo_microwave-based_2023}%
  \BibitemOpen
  \bibfield  {author} {\bibinfo {author} {\bibfnamefont {X.}~\bibnamefont {Guo}}, \bibinfo {author} {\bibfnamefont {A.~M.}\ \bibnamefont {Stramma}}, \bibinfo {author} {\bibfnamefont {Z.}~\bibnamefont {Li}}, \bibinfo {author} {\bibfnamefont {W.~G.}\ \bibnamefont {Roth}}, \bibinfo {author} {\bibfnamefont {B.}~\bibnamefont {Huang}}, \bibinfo {author} {\bibfnamefont {Y.}~\bibnamefont {Jin}}, \bibinfo {author} {\bibfnamefont {R.~A.}\ \bibnamefont {Parker}}, \bibinfo {author} {\bibfnamefont {J.}~\bibnamefont {Arjona~Mart{\'\i}nez}}, \bibinfo {author} {\bibfnamefont {N.}~\bibnamefont {Shofer}}, \bibinfo {author} {\bibfnamefont {C.~P.}\ \bibnamefont {Michaels}}, \bibinfo {author} {\bibfnamefont {C.~P.}\ \bibnamefont {Purser}}, \bibinfo {author} {\bibfnamefont {M.~H.}\ \bibnamefont {Appel}}, \bibinfo {author} {\bibfnamefont {E.~M.}\ \bibnamefont {Alexeev}}, \bibinfo {author} {\bibfnamefont {T.}~\bibnamefont {Liu}}, \bibinfo {author} {\bibfnamefont {A.~C.}\ \bibnamefont {Ferrari}}, \bibinfo {author} {\bibfnamefont
  {D.~D.}\ \bibnamefont {Awschalom}}, \bibinfo {author} {\bibfnamefont {N.}~\bibnamefont {Delegan}}, \bibinfo {author} {\bibfnamefont {B.}~\bibnamefont {Pingault}}, \bibinfo {author} {\bibfnamefont {G.}~\bibnamefont {Galli}}, \bibinfo {author} {\bibfnamefont {F.~J.}\ \bibnamefont {Heremans}}, \bibinfo {author} {\bibfnamefont {M.}~\bibnamefont {Atat{\"u}re}},\ and\ \bibinfo {author} {\bibfnamefont {A.~A.}\ \bibnamefont {High}},\ }\bibfield  {title} {\bibinfo {title} {Microwave-{Based} {Quantum} {Control} and {Coherence} {Protection} of {Tin}-{Vacancy} {Spin} {Qubits} in a {Strain}-{Tuned} {Diamond}-{Membrane} {Heterostructure}},\ }\href {https://doi.org/10.1103/PhysRevX.13.041037} {\bibfield  {journal} {\bibinfo  {journal} {Physical Review X}\ }\textbf {\bibinfo {volume} {13}},\ \bibinfo {pages} {041037} (\bibinfo {year} {2023})}\BibitemShut {NoStop}%
\bibitem [{\citenamefont {Rosenthal}\ \emph {et~al.}(2023)\citenamefont {Rosenthal}, \citenamefont {Anderson}, \citenamefont {Kleidermacher}, \citenamefont {Stein}, \citenamefont {Lee}, \citenamefont {Grzesik}, \citenamefont {Scuri}, \citenamefont {Rugar}, \citenamefont {Riedel}, \citenamefont {Aghaeimeibodi}, \citenamefont {Ahn}, \citenamefont {Van~Gasse},\ and\ \citenamefont {Vu{\v c}kovi{\'c}}}]{rosenthal_microwave_2023}%
  \BibitemOpen
  \bibfield  {author} {\bibinfo {author} {\bibfnamefont {E.~I.}\ \bibnamefont {Rosenthal}}, \bibinfo {author} {\bibfnamefont {C.~P.}\ \bibnamefont {Anderson}}, \bibinfo {author} {\bibfnamefont {H.~C.}\ \bibnamefont {Kleidermacher}}, \bibinfo {author} {\bibfnamefont {A.~J.}\ \bibnamefont {Stein}}, \bibinfo {author} {\bibfnamefont {H.}~\bibnamefont {Lee}}, \bibinfo {author} {\bibfnamefont {J.}~\bibnamefont {Grzesik}}, \bibinfo {author} {\bibfnamefont {G.}~\bibnamefont {Scuri}}, \bibinfo {author} {\bibfnamefont {A.~E.}\ \bibnamefont {Rugar}}, \bibinfo {author} {\bibfnamefont {D.}~\bibnamefont {Riedel}}, \bibinfo {author} {\bibfnamefont {S.}~\bibnamefont {Aghaeimeibodi}}, \bibinfo {author} {\bibfnamefont {G.~H.}\ \bibnamefont {Ahn}}, \bibinfo {author} {\bibfnamefont {K.}~\bibnamefont {Van~Gasse}},\ and\ \bibinfo {author} {\bibfnamefont {J.}~\bibnamefont {Vu{\v c}kovi{\'c}}},\ }\bibfield  {title} {\bibinfo {title} {Microwave {Spin} {Control} of a {Tin}-{Vacancy} {Qubit} in {Diamond}},\ }\href
  {https://doi.org/10.1103/PhysRevX.13.031022} {\bibfield  {journal} {\bibinfo  {journal} {Physical Review X}\ }\textbf {\bibinfo {volume} {13}},\ \bibinfo {pages} {031022} (\bibinfo {year} {2023})}\BibitemShut {NoStop}%
\bibitem [{\citenamefont {Karapatzakis}\ \emph {et~al.}(2024)\citenamefont {Karapatzakis}, \citenamefont {Resch}, \citenamefont {Schrodin}, \citenamefont {Fuchs}, \citenamefont {Kieschnick}, \citenamefont {Heupel}, \citenamefont {Kussi}, \citenamefont {S{\"u}rgers}, \citenamefont {Popov}, \citenamefont {Meijer}, \citenamefont {Becher}, \citenamefont {Wernsdorfer},\ and\ \citenamefont {Hunger}}]{karapatzakis_microwave_2024}%
  \BibitemOpen
  \bibfield  {author} {\bibinfo {author} {\bibfnamefont {I.}~\bibnamefont {Karapatzakis}}, \bibinfo {author} {\bibfnamefont {J.}~\bibnamefont {Resch}}, \bibinfo {author} {\bibfnamefont {M.}~\bibnamefont {Schrodin}}, \bibinfo {author} {\bibfnamefont {P.}~\bibnamefont {Fuchs}}, \bibinfo {author} {\bibfnamefont {M.}~\bibnamefont {Kieschnick}}, \bibinfo {author} {\bibfnamefont {J.}~\bibnamefont {Heupel}}, \bibinfo {author} {\bibfnamefont {L.}~\bibnamefont {Kussi}}, \bibinfo {author} {\bibfnamefont {C.}~\bibnamefont {S{\"u}rgers}}, \bibinfo {author} {\bibfnamefont {C.}~\bibnamefont {Popov}}, \bibinfo {author} {\bibfnamefont {J.}~\bibnamefont {Meijer}}, \bibinfo {author} {\bibfnamefont {C.}~\bibnamefont {Becher}}, \bibinfo {author} {\bibfnamefont {W.}~\bibnamefont {Wernsdorfer}},\ and\ \bibinfo {author} {\bibfnamefont {D.}~\bibnamefont {Hunger}},\ }\bibfield  {title} {\bibinfo {title} {Microwave {Control} of the {Tin}-{Vacancy} {Spin} {Qubit} in {Diamond} with a {Superconducting} {Waveguide}},\ }\href
  {https://doi.org/10.1103/PhysRevX.14.031036} {\bibfield  {journal} {\bibinfo  {journal} {Physical Review X}\ }\textbf {\bibinfo {volume} {14}},\ \bibinfo {pages} {031036} (\bibinfo {year} {2024})}\BibitemShut {NoStop}%
\bibitem [{\citenamefont {Li}\ \emph {et~al.}(2024{\natexlab{a}})\citenamefont {Li}, \citenamefont {Santis}, \citenamefont {Harris}, \citenamefont {Chen}, \citenamefont {Gao}, \citenamefont {Christen}, \citenamefont {Choi}, \citenamefont {Trusheim}, \citenamefont {Song}, \citenamefont {Errando-Herranz}, \citenamefont {Du}, \citenamefont {Hu}, \citenamefont {Clark}, \citenamefont {Ibrahim}, \citenamefont {Gilbert}, \citenamefont {Han},\ and\ \citenamefont {Englund}}]{li_heterogeneous_2024}%
  \BibitemOpen
  \bibfield  {author} {\bibinfo {author} {\bibfnamefont {L.}~\bibnamefont {Li}}, \bibinfo {author} {\bibfnamefont {L.~D.}\ \bibnamefont {Santis}}, \bibinfo {author} {\bibfnamefont {I.~B.~W.}\ \bibnamefont {Harris}}, \bibinfo {author} {\bibfnamefont {K.~C.}\ \bibnamefont {Chen}}, \bibinfo {author} {\bibfnamefont {Y.}~\bibnamefont {Gao}}, \bibinfo {author} {\bibfnamefont {I.}~\bibnamefont {Christen}}, \bibinfo {author} {\bibfnamefont {H.}~\bibnamefont {Choi}}, \bibinfo {author} {\bibfnamefont {M.}~\bibnamefont {Trusheim}}, \bibinfo {author} {\bibfnamefont {Y.}~\bibnamefont {Song}}, \bibinfo {author} {\bibfnamefont {C.}~\bibnamefont {Errando-Herranz}}, \bibinfo {author} {\bibfnamefont {J.}~\bibnamefont {Du}}, \bibinfo {author} {\bibfnamefont {Y.}~\bibnamefont {Hu}}, \bibinfo {author} {\bibfnamefont {G.}~\bibnamefont {Clark}}, \bibinfo {author} {\bibfnamefont {M.~I.}\ \bibnamefont {Ibrahim}}, \bibinfo {author} {\bibfnamefont {G.}~\bibnamefont {Gilbert}}, \bibinfo {author} {\bibfnamefont {R.}~\bibnamefont {Han}},\
  and\ \bibinfo {author} {\bibfnamefont {D.}~\bibnamefont {Englund}},\ }\bibfield  {title} {\bibinfo {title} {Heterogeneous integration of spin--photon interfaces with a {CMOS} platform},\ }\href {https://doi.org/10.1038/s41586-024-07371-7} {\bibfield  {journal} {\bibinfo  {journal} {Nature}\ }\textbf {\bibinfo {volume} {630}},\ \bibinfo {pages} {70} (\bibinfo {year} {2024}{\natexlab{a}})},\ \bibinfo {note} {publisher: Nature Publishing Group}\BibitemShut {NoStop}%
\bibitem [{\citenamefont {Narita}\ \emph {et~al.}(2023)\citenamefont {Narita}, \citenamefont {Wang}, \citenamefont {Ikeda}, \citenamefont {Oba}, \citenamefont {Miyamoto}, \citenamefont {Taniguchi}, \citenamefont {Onoda}, \citenamefont {Hatano},\ and\ \citenamefont {Iwasaki}}]{narita_multiple_2023}%
  \BibitemOpen
  \bibfield  {author} {\bibinfo {author} {\bibfnamefont {Y.}~\bibnamefont {Narita}}, \bibinfo {author} {\bibfnamefont {P.}~\bibnamefont {Wang}}, \bibinfo {author} {\bibfnamefont {K.}~\bibnamefont {Ikeda}}, \bibinfo {author} {\bibfnamefont {K.}~\bibnamefont {Oba}}, \bibinfo {author} {\bibfnamefont {Y.}~\bibnamefont {Miyamoto}}, \bibinfo {author} {\bibfnamefont {T.}~\bibnamefont {Taniguchi}}, \bibinfo {author} {\bibfnamefont {S.}~\bibnamefont {Onoda}}, \bibinfo {author} {\bibfnamefont {M.}~\bibnamefont {Hatano}},\ and\ \bibinfo {author} {\bibfnamefont {T.}~\bibnamefont {Iwasaki}},\ }\bibfield  {title} {\bibinfo {title} {Multiple {Tin}-{Vacancy} {Centers} in {Diamond} with {Nearly} {Identical} {Photon} {Frequency} and {Linewidth}},\ }\href {https://doi.org/10.1103/PhysRevApplied.19.024061} {\bibfield  {journal} {\bibinfo  {journal} {Physical Review Applied}\ }\textbf {\bibinfo {volume} {19}},\ \bibinfo {pages} {024061} (\bibinfo {year} {2023})}\BibitemShut {NoStop}%
\bibitem [{\citenamefont {Aghaeimeibodi}\ \emph {et~al.}(2021)\citenamefont {Aghaeimeibodi}, \citenamefont {Riedel}, \citenamefont {Rugar}, \citenamefont {Dory},\ and\ \citenamefont {Vu{\v c}kovi{\'c}}}]{aghaeimeibodi_electrical_2021}%
  \BibitemOpen
  \bibfield  {author} {\bibinfo {author} {\bibfnamefont {S.}~\bibnamefont {Aghaeimeibodi}}, \bibinfo {author} {\bibfnamefont {D.}~\bibnamefont {Riedel}}, \bibinfo {author} {\bibfnamefont {A.~E.}\ \bibnamefont {Rugar}}, \bibinfo {author} {\bibfnamefont {C.}~\bibnamefont {Dory}},\ and\ \bibinfo {author} {\bibfnamefont {J.}~\bibnamefont {Vu{\v c}kovi{\'c}}},\ }\bibfield  {title} {\bibinfo {title} {Electrical {Tuning} of {Tin}-{Vacancy} {Centers} in {Diamond}},\ }\href {https://doi.org/10.1103/PhysRevApplied.15.064010} {\bibfield  {journal} {\bibinfo  {journal} {Physical Review Applied}\ }\textbf {\bibinfo {volume} {15}},\ \bibinfo {pages} {064010} (\bibinfo {year} {2021})}\BibitemShut {NoStop}%
\bibitem [{\citenamefont {De~Santis}\ \emph {et~al.}(2021)\citenamefont {De~Santis}, \citenamefont {Trusheim}, \citenamefont {Chen},\ and\ \citenamefont {Englund}}]{de_santis_investigation_2021}%
  \BibitemOpen
  \bibfield  {author} {\bibinfo {author} {\bibfnamefont {L.}~\bibnamefont {De~Santis}}, \bibinfo {author} {\bibfnamefont {M.~E.}\ \bibnamefont {Trusheim}}, \bibinfo {author} {\bibfnamefont {K.~C.}\ \bibnamefont {Chen}},\ and\ \bibinfo {author} {\bibfnamefont {D.~R.}\ \bibnamefont {Englund}},\ }\bibfield  {title} {\bibinfo {title} {Investigation of the {Stark} {Effect} on a {Centrosymmetric} {Quantum} {Emitter} in {Diamond}},\ }\href {https://doi.org/10.1103/PhysRevLett.127.147402} {\bibfield  {journal} {\bibinfo  {journal} {Physical Review Letters}\ }\textbf {\bibinfo {volume} {127}},\ \bibinfo {pages} {147402} (\bibinfo {year} {2021})}\BibitemShut {NoStop}%
\bibitem [{\citenamefont {Lehtinen}\ \emph {et~al.}(2016)\citenamefont {Lehtinen}, \citenamefont {Naydenov}, \citenamefont {B{\"o}rner}, \citenamefont {Melentjevic}, \citenamefont {M{\"u}ller}, \citenamefont {McGuinness}, \citenamefont {Pezzagna}, \citenamefont {Meijer}, \citenamefont {Kaiser},\ and\ \citenamefont {Jelezko}}]{lehtinen_molecular_2016}%
  \BibitemOpen
  \bibfield  {author} {\bibinfo {author} {\bibfnamefont {O.}~\bibnamefont {Lehtinen}}, \bibinfo {author} {\bibfnamefont {B.}~\bibnamefont {Naydenov}}, \bibinfo {author} {\bibfnamefont {P.}~\bibnamefont {B{\"o}rner}}, \bibinfo {author} {\bibfnamefont {K.}~\bibnamefont {Melentjevic}}, \bibinfo {author} {\bibfnamefont {C.}~\bibnamefont {M{\"u}ller}}, \bibinfo {author} {\bibfnamefont {L.~P.}\ \bibnamefont {McGuinness}}, \bibinfo {author} {\bibfnamefont {S.}~\bibnamefont {Pezzagna}}, \bibinfo {author} {\bibfnamefont {J.}~\bibnamefont {Meijer}}, \bibinfo {author} {\bibfnamefont {U.}~\bibnamefont {Kaiser}},\ and\ \bibinfo {author} {\bibfnamefont {F.}~\bibnamefont {Jelezko}},\ }\bibfield  {title} {\bibinfo {title} {Molecular dynamics simulations of shallow nitrogen and silicon implantation into diamond},\ }\href {https://doi.org/10.1103/PhysRevB.93.035202} {\bibfield  {journal} {\bibinfo  {journal} {Physical Review B}\ }\textbf {\bibinfo {volume} {93}},\ \bibinfo {pages} {035202} (\bibinfo {year} {2016})}\BibitemShut
  {NoStop}%
\bibitem [{\citenamefont {Santonocito}\ \emph {et~al.}(2024)\citenamefont {Santonocito}, \citenamefont {Denisenko}, \citenamefont {Schreck}, \citenamefont {Pasquarelli},\ and\ \citenamefont {Wrachtrup}}]{santonocito_suppression_2024}%
  \BibitemOpen
  \bibfield  {author} {\bibinfo {author} {\bibfnamefont {S.}~\bibnamefont {Santonocito}}, \bibinfo {author} {\bibfnamefont {A.}~\bibnamefont {Denisenko}}, \bibinfo {author} {\bibfnamefont {M.}~\bibnamefont {Schreck}}, \bibinfo {author} {\bibfnamefont {A.}~\bibnamefont {Pasquarelli}},\ and\ \bibinfo {author} {\bibfnamefont {J.}~\bibnamefont {Wrachtrup}},\ }\bibfield  {title} {\bibinfo {title} {Suppression of thermal diffusion of vacancies across p-n junction structures in diamond. {Application} to {SnV} centers by ion implantation},\ }\href {https://doi.org/10.1088/1367-2630/ad44cd} {\bibfield  {journal} {\bibinfo  {journal} {New Journal of Physics}\ }\textbf {\bibinfo {volume} {26}},\ \bibinfo {pages} {053036} (\bibinfo {year} {2024})}\BibitemShut {NoStop}%
\bibitem [{\citenamefont {Wang}\ \emph {et~al.}(2022)\citenamefont {Wang}, \citenamefont {Bushmakin}, \citenamefont {Stein}, \citenamefont {Schell},\ and\ \citenamefont {Gerhardt}}]{wang_optical_2022}%
  \BibitemOpen
  \bibfield  {author} {\bibinfo {author} {\bibfnamefont {Y.}~\bibnamefont {Wang}}, \bibinfo {author} {\bibfnamefont {V.}~\bibnamefont {Bushmakin}}, \bibinfo {author} {\bibfnamefont {G.~A.}\ \bibnamefont {Stein}}, \bibinfo {author} {\bibfnamefont {A.~W.}\ \bibnamefont {Schell}},\ and\ \bibinfo {author} {\bibfnamefont {I.}~\bibnamefont {Gerhardt}},\ }\bibfield  {title} {\bibinfo {title} {Optical {Ramsey} spectroscopy on a single molecule},\ }\href {https://doi.org/10.1364/OPTICA.443727} {\bibfield  {journal} {\bibinfo  {journal} {Optica}\ }\textbf {\bibinfo {volume} {9}},\ \bibinfo {pages} {374} (\bibinfo {year} {2022})}\BibitemShut {NoStop}%
\bibitem [{\citenamefont {Segura}\ \emph {et~al.}(2001)\citenamefont {Segura}, \citenamefont {Zumofen}, \citenamefont {Renn}, \citenamefont {Hecht},\ and\ \citenamefont {Wild}}]{segura_tip-induced_2001}%
  \BibitemOpen
  \bibfield  {author} {\bibinfo {author} {\bibfnamefont {J.~M.}\ \bibnamefont {Segura}}, \bibinfo {author} {\bibfnamefont {G.}~\bibnamefont {Zumofen}}, \bibinfo {author} {\bibfnamefont {A.}~\bibnamefont {Renn}}, \bibinfo {author} {\bibfnamefont {B.}~\bibnamefont {Hecht}},\ and\ \bibinfo {author} {\bibfnamefont {U.~P.}\ \bibnamefont {Wild}},\ }\bibfield  {title} {\bibinfo {title} {Tip-induced spectral dynamics of single molecules},\ }\href {https://doi.org/10.1016/S0009-2614(01)00384-0} {\bibfield  {journal} {\bibinfo  {journal} {Chemical Physics Letters}\ }\textbf {\bibinfo {volume} {340}},\ \bibinfo {pages} {77} (\bibinfo {year} {2001})}\BibitemShut {NoStop}%
\bibitem [{\citenamefont {Bassett}\ \emph {et~al.}(2011)\citenamefont {Bassett}, \citenamefont {Heremans}, \citenamefont {Yale}, \citenamefont {Buckley},\ and\ \citenamefont {Awschalom}}]{bassett_electrical_2011}%
  \BibitemOpen
  \bibfield  {author} {\bibinfo {author} {\bibfnamefont {L.~C.}\ \bibnamefont {Bassett}}, \bibinfo {author} {\bibfnamefont {F.~J.}\ \bibnamefont {Heremans}}, \bibinfo {author} {\bibfnamefont {C.~G.}\ \bibnamefont {Yale}}, \bibinfo {author} {\bibfnamefont {B.~B.}\ \bibnamefont {Buckley}},\ and\ \bibinfo {author} {\bibfnamefont {D.~D.}\ \bibnamefont {Awschalom}},\ }\bibfield  {title} {\bibinfo {title} {Electrical {Tuning} of {Single} {Nitrogen}-{Vacancy} {Center} {Optical} {Transitions} {Enhanced} by {Photoinduced} {Fields}},\ }\href {https://doi.org/10.1103/PhysRevLett.107.266403} {\bibfield  {journal} {\bibinfo  {journal} {Physical Review Letters}\ }\textbf {\bibinfo {volume} {107}},\ \bibinfo {pages} {266403} (\bibinfo {year} {2011})}\BibitemShut {NoStop}%
\bibitem [{\citenamefont {G{\"o}rlitz}\ \emph {et~al.}(2020)\citenamefont {G{\"o}rlitz}, \citenamefont {Herrmann}, \citenamefont {Thiering}, \citenamefont {Fuchs}, \citenamefont {Gandil}, \citenamefont {Iwasaki}, \citenamefont {Taniguchi}, \citenamefont {Kieschnick}, \citenamefont {Meijer}, \citenamefont {Hatano}, \citenamefont {Gali},\ and\ \citenamefont {Becher}}]{gorlitz_spectroscopic_2020}%
  \BibitemOpen
  \bibfield  {author} {\bibinfo {author} {\bibfnamefont {J.}~\bibnamefont {G{\"o}rlitz}}, \bibinfo {author} {\bibfnamefont {D.}~\bibnamefont {Herrmann}}, \bibinfo {author} {\bibfnamefont {G.}~\bibnamefont {Thiering}}, \bibinfo {author} {\bibfnamefont {P.}~\bibnamefont {Fuchs}}, \bibinfo {author} {\bibfnamefont {M.}~\bibnamefont {Gandil}}, \bibinfo {author} {\bibfnamefont {T.}~\bibnamefont {Iwasaki}}, \bibinfo {author} {\bibfnamefont {T.}~\bibnamefont {Taniguchi}}, \bibinfo {author} {\bibfnamefont {M.}~\bibnamefont {Kieschnick}}, \bibinfo {author} {\bibfnamefont {J.}~\bibnamefont {Meijer}}, \bibinfo {author} {\bibfnamefont {M.}~\bibnamefont {Hatano}}, \bibinfo {author} {\bibfnamefont {A.}~\bibnamefont {Gali}},\ and\ \bibinfo {author} {\bibfnamefont {C.}~\bibnamefont {Becher}},\ }\bibfield  {title} {\bibinfo {title} {Spectroscopic investigations of negatively charged tin-vacancy centres in diamond},\ }\href {https://doi.org/10.1088/1367-2630/ab6631} {\bibfield  {journal} {\bibinfo  {journal} {New Journal of
  Physics}\ }\textbf {\bibinfo {volume} {22}},\ \bibinfo {pages} {013048} (\bibinfo {year} {2020})}\BibitemShut {NoStop}%
\bibitem [{\citenamefont {Wang}\ \emph {et~al.}(2024)\citenamefont {Wang}, \citenamefont {Kazak}, \citenamefont {Senkalla}, \citenamefont {Siyushev}, \citenamefont {Abe}, \citenamefont {Taniguchi}, \citenamefont {Onoda}, \citenamefont {Kato}, \citenamefont {Makino}, \citenamefont {Hatano}, \citenamefont {Jelezko},\ and\ \citenamefont {Iwasaki}}]{wang_transform-limited_2024}%
  \BibitemOpen
  \bibfield  {author} {\bibinfo {author} {\bibfnamefont {P.}~\bibnamefont {Wang}}, \bibinfo {author} {\bibfnamefont {L.}~\bibnamefont {Kazak}}, \bibinfo {author} {\bibfnamefont {K.}~\bibnamefont {Senkalla}}, \bibinfo {author} {\bibfnamefont {P.}~\bibnamefont {Siyushev}}, \bibinfo {author} {\bibfnamefont {R.}~\bibnamefont {Abe}}, \bibinfo {author} {\bibfnamefont {T.}~\bibnamefont {Taniguchi}}, \bibinfo {author} {\bibfnamefont {S.}~\bibnamefont {Onoda}}, \bibinfo {author} {\bibfnamefont {H.}~\bibnamefont {Kato}}, \bibinfo {author} {\bibfnamefont {T.}~\bibnamefont {Makino}}, \bibinfo {author} {\bibfnamefont {M.}~\bibnamefont {Hatano}}, \bibinfo {author} {\bibfnamefont {F.}~\bibnamefont {Jelezko}},\ and\ \bibinfo {author} {\bibfnamefont {T.}~\bibnamefont {Iwasaki}},\ }\bibfield  {title} {\bibinfo {title} {Transform-{Limited} {Photon} {Emission} from a {Lead}-{Vacancy} {Center} in {Diamond} above 10 {K}},\ }\href {https://doi.org/10.1103/PhysRevLett.132.073601} {\bibfield  {journal} {\bibinfo  {journal} {Physical
  Review Letters}\ }\textbf {\bibinfo {volume} {132}},\ \bibinfo {pages} {073601} (\bibinfo {year} {2024})}\BibitemShut {NoStop}%
\bibitem [{\citenamefont {Hong}\ \emph {et~al.}()\citenamefont {Hong}, \citenamefont {Ou},\ and\ \citenamefont {Mandel}}]{hong_measurement_1987}%
  \BibitemOpen
  \bibfield  {author} {\bibinfo {author} {\bibfnamefont {C.~K.}\ \bibnamefont {Hong}}, \bibinfo {author} {\bibfnamefont {Z.~Y.}\ \bibnamefont {Ou}},\ and\ \bibinfo {author} {\bibfnamefont {L.}~\bibnamefont {Mandel}},\ }\bibfield  {title} {\bibinfo {title} {Measurement of subpicosecond time intervals between two photons by interference},\ }\href {https://doi.org/10.1103/PhysRevLett.59.2044} {\bibfield  {journal} {\bibinfo  {journal} {Physical Review Letters}\ }\textbf {\bibinfo {volume} {59}},\ \bibinfo {pages} {2044}}\BibitemShut {NoStop}%
\bibitem [{\citenamefont {Shih}\ and\ \citenamefont {Alley}()}]{shih_new_1988}%
  \BibitemOpen
  \bibfield  {author} {\bibinfo {author} {\bibfnamefont {Y.~H.}\ \bibnamefont {Shih}}\ and\ \bibinfo {author} {\bibfnamefont {C.~O.}\ \bibnamefont {Alley}},\ }\bibfield  {title} {\bibinfo {title} {New type of einstein-podolsky-rosen-bohm experiment using pairs of light quanta produced by optical parametric down conversion},\ }\href {https://doi.org/10.1103/PhysRevLett.61.2921} {\bibfield  {journal} {\bibinfo  {journal} {Physical Review Letters}\ }\textbf {\bibinfo {volume} {61}},\ \bibinfo {pages} {2921}}\BibitemShut {NoStop}%
\bibitem [{\citenamefont {Morioka}\ \emph {et~al.}(2020{\natexlab{b}})\citenamefont {Morioka}, \citenamefont {Babin}, \citenamefont {Nagy}, \citenamefont {Gediz}, \citenamefont {Hesselmeier}, \citenamefont {Liu}, \citenamefont {Joliffe}, \citenamefont {Niethammer}, \citenamefont {Dasari}, \citenamefont {Vorobyov}, \citenamefont {Kolesov}, \citenamefont {St{\"o}hr}, \citenamefont {Ul-Hassan}, \citenamefont {Son}, \citenamefont {Ohshima}, \citenamefont {Udvarhelyi}, \citenamefont {Thiering}, \citenamefont {Gali}, \citenamefont {Wrachtrup},\ and\ \citenamefont {Kaiser}}]{morioka_spin-controlled_2020}%
  \BibitemOpen
  \bibfield  {author} {\bibinfo {author} {\bibfnamefont {N.}~\bibnamefont {Morioka}}, \bibinfo {author} {\bibfnamefont {C.}~\bibnamefont {Babin}}, \bibinfo {author} {\bibfnamefont {R.}~\bibnamefont {Nagy}}, \bibinfo {author} {\bibfnamefont {I.}~\bibnamefont {Gediz}}, \bibinfo {author} {\bibfnamefont {E.}~\bibnamefont {Hesselmeier}}, \bibinfo {author} {\bibfnamefont {D.}~\bibnamefont {Liu}}, \bibinfo {author} {\bibfnamefont {M.}~\bibnamefont {Joliffe}}, \bibinfo {author} {\bibfnamefont {M.}~\bibnamefont {Niethammer}}, \bibinfo {author} {\bibfnamefont {D.}~\bibnamefont {Dasari}}, \bibinfo {author} {\bibfnamefont {V.}~\bibnamefont {Vorobyov}}, \bibinfo {author} {\bibfnamefont {R.}~\bibnamefont {Kolesov}}, \bibinfo {author} {\bibfnamefont {R.}~\bibnamefont {St{\"o}hr}}, \bibinfo {author} {\bibfnamefont {J.}~\bibnamefont {Ul-Hassan}}, \bibinfo {author} {\bibfnamefont {N.~T.}\ \bibnamefont {Son}}, \bibinfo {author} {\bibfnamefont {T.}~\bibnamefont {Ohshima}}, \bibinfo {author} {\bibfnamefont {P.}~\bibnamefont
  {Udvarhelyi}}, \bibinfo {author} {\bibfnamefont {G.}~\bibnamefont {Thiering}}, \bibinfo {author} {\bibfnamefont {A.}~\bibnamefont {Gali}}, \bibinfo {author} {\bibfnamefont {J.}~\bibnamefont {Wrachtrup}},\ and\ \bibinfo {author} {\bibfnamefont {F.}~\bibnamefont {Kaiser}},\ }\bibfield  {title} {\bibinfo {title} {Spin-controlled generation of indistinguishable and distinguishable photons from silicon vacancy centres in silicon carbide},\ }\href {https://doi.org/10.1038/s41467-020-16330-5} {\bibfield  {journal} {\bibinfo  {journal} {Nature Communications}\ }\textbf {\bibinfo {volume} {11}},\ \bibinfo {pages} {2516} (\bibinfo {year} {2020}{\natexlab{b}})}\BibitemShut {NoStop}%
\bibitem [{\citenamefont {Rezai}\ \emph {et~al.}(2018)\citenamefont {Rezai}, \citenamefont {Wrachtrup},\ and\ \citenamefont {Gerhardt}}]{rezai_coherence_2018}%
  \BibitemOpen
  \bibfield  {author} {\bibinfo {author} {\bibfnamefont {M.}~\bibnamefont {Rezai}}, \bibinfo {author} {\bibfnamefont {J.}~\bibnamefont {Wrachtrup}},\ and\ \bibinfo {author} {\bibfnamefont {I.}~\bibnamefont {Gerhardt}},\ }\bibfield  {title} {\bibinfo {title} {Coherence {Properties} of {Molecular} {Single} {Photons} for {Quantum} {Networks}},\ }\href {https://doi.org/10.1103/PhysRevX.8.031026} {\bibfield  {journal} {\bibinfo  {journal} {Physical Review X}\ }\textbf {\bibinfo {volume} {8}},\ \bibinfo {pages} {031026} (\bibinfo {year} {2018})}\BibitemShut {NoStop}%
\bibitem [{\citenamefont {Legero}\ \emph {et~al.}(2004)\citenamefont {Legero}, \citenamefont {Wilk}, \citenamefont {Hennrich}, \citenamefont {Rempe},\ and\ \citenamefont {Kuhn}}]{legero_quantum_2004}%
  \BibitemOpen
  \bibfield  {author} {\bibinfo {author} {\bibfnamefont {T.}~\bibnamefont {Legero}}, \bibinfo {author} {\bibfnamefont {T.}~\bibnamefont {Wilk}}, \bibinfo {author} {\bibfnamefont {M.}~\bibnamefont {Hennrich}}, \bibinfo {author} {\bibfnamefont {G.}~\bibnamefont {Rempe}},\ and\ \bibinfo {author} {\bibfnamefont {A.}~\bibnamefont {Kuhn}},\ }\bibfield  {title} {\bibinfo {title} {Quantum {Beat} of {Two} {Single} {Photons}},\ }\href {https://doi.org/10.1103/PhysRevLett.93.070503} {\bibfield  {journal} {\bibinfo  {journal} {Physical Review Letters}\ }\textbf {\bibinfo {volume} {93}},\ \bibinfo {pages} {070503} (\bibinfo {year} {2004})}\BibitemShut {NoStop}%
\bibitem [{\citenamefont {Lettow}\ \emph {et~al.}(2010)\citenamefont {Lettow}, \citenamefont {Rezus}, \citenamefont {Renn}, \citenamefont {Zumofen}, \citenamefont {Ikonen}, \citenamefont {G{\"o}tzinger},\ and\ \citenamefont {Sandoghdar}}]{lettow_quantum_2010}%
  \BibitemOpen
  \bibfield  {author} {\bibinfo {author} {\bibfnamefont {R.}~\bibnamefont {Lettow}}, \bibinfo {author} {\bibfnamefont {Y.~L.~A.}\ \bibnamefont {Rezus}}, \bibinfo {author} {\bibfnamefont {A.}~\bibnamefont {Renn}}, \bibinfo {author} {\bibfnamefont {G.}~\bibnamefont {Zumofen}}, \bibinfo {author} {\bibfnamefont {E.}~\bibnamefont {Ikonen}}, \bibinfo {author} {\bibfnamefont {S.}~\bibnamefont {G{\"o}tzinger}},\ and\ \bibinfo {author} {\bibfnamefont {V.}~\bibnamefont {Sandoghdar}},\ }\bibfield  {title} {\bibinfo {title} {Quantum {Interference} of {Tunably} {Indistinguishable} {Photons} from {Remote} {Organic} {Molecules}},\ }\href {https://doi.org/10.1103/PhysRevLett.104.123605} {\bibfield  {journal} {\bibinfo  {journal} {Physical Review Letters}\ }\textbf {\bibinfo {volume} {104}},\ \bibinfo {pages} {123605} (\bibinfo {year} {2010})},\ \bibinfo {note} {publisher: American Physical Society}\BibitemShut {NoStop}%
\bibitem [{\citenamefont {Li}\ \emph {et~al.}(2024{\natexlab{b}})\citenamefont {Li}, \citenamefont {Guo}, \citenamefont {Jin}, \citenamefont {Andreoli}, \citenamefont {Bilgin}, \citenamefont {Awschalom}, \citenamefont {Delegan}, \citenamefont {Heremans}, \citenamefont {Chang}, \citenamefont {Galli},\ and\ \citenamefont {High}}]{li_atomic_2024}%
  \BibitemOpen
  \bibfield  {author} {\bibinfo {author} {\bibfnamefont {Z.}~\bibnamefont {Li}}, \bibinfo {author} {\bibfnamefont {X.}~\bibnamefont {Guo}}, \bibinfo {author} {\bibfnamefont {Y.}~\bibnamefont {Jin}}, \bibinfo {author} {\bibfnamefont {F.}~\bibnamefont {Andreoli}}, \bibinfo {author} {\bibfnamefont {A.}~\bibnamefont {Bilgin}}, \bibinfo {author} {\bibfnamefont {D.~D.}\ \bibnamefont {Awschalom}}, \bibinfo {author} {\bibfnamefont {N.}~\bibnamefont {Delegan}}, \bibinfo {author} {\bibfnamefont {F.~J.}\ \bibnamefont {Heremans}}, \bibinfo {author} {\bibfnamefont {D.}~\bibnamefont {Chang}}, \bibinfo {author} {\bibfnamefont {G.}~\bibnamefont {Galli}},\ and\ \bibinfo {author} {\bibfnamefont {A.~A.}\ \bibnamefont {High}},\ }\bibfield  {title} {\bibinfo {title} {{Atomic optical antennas in solids}},\ }\href {https://doi.org/10.1038/s41566-024-01456-5} {\bibfield  {journal} {\bibinfo  {journal} {Nature Photonics}\ }\textbf {\bibinfo {volume} {18}},\ \bibinfo {pages} {1113} (\bibinfo {year} {2024}{\natexlab{b}})}\BibitemShut
  {NoStop}%
\bibitem [{\citenamefont {Pieplow}\ \emph {et~al.}(2024)\citenamefont {Pieplow}, \citenamefont {Torun}, \citenamefont {Munns}, \citenamefont {Herrmann}, \citenamefont {Thies}, \citenamefont {Pregnolato},\ and\ \citenamefont {Schr{\"o}der}}]{pieplow_quantum_2024}%
  \BibitemOpen
  \bibfield  {author} {\bibinfo {author} {\bibfnamefont {G.}~\bibnamefont {Pieplow}}, \bibinfo {author} {\bibfnamefont {C.~G.}\ \bibnamefont {Torun}}, \bibinfo {author} {\bibfnamefont {J.~H.~D.}\ \bibnamefont {Munns}}, \bibinfo {author} {\bibfnamefont {F.~M.}\ \bibnamefont {Herrmann}}, \bibinfo {author} {\bibfnamefont {A.}~\bibnamefont {Thies}}, \bibinfo {author} {\bibfnamefont {T.}~\bibnamefont {Pregnolato}},\ and\ \bibinfo {author} {\bibfnamefont {T.}~\bibnamefont {Schr{\"o}der}},\ }\href {https://doi.org/10.48550/arXiv.2401.14290} {\bibinfo {title} {Quantum {Electrometer} for {Time}-{Resolved} {Material} {Science} at the {Atomic} {Lattice} {Scale}}} (\bibinfo {year} {2024}),\ \bibinfo {note} {arXiv:2401.14290}\BibitemShut {NoStop}%
\bibitem [{\citenamefont {Machielse}\ \emph {et~al.}(2019)\citenamefont {Machielse}, \citenamefont {Bogdanovic}, \citenamefont {Meesala}, \citenamefont {Gauthier}, \citenamefont {Burek}, \citenamefont {Joe}, \citenamefont {Chalupnik}, \citenamefont {Sohn}, \citenamefont {Holzgrafe}, \citenamefont {Evans}, \citenamefont {Chia}, \citenamefont {Atikian}, \citenamefont {Bhaskar}, \citenamefont {Sukachev}, \citenamefont {Shao}, \citenamefont {Maity}, \citenamefont {Lukin},\ and\ \citenamefont {Lon{\v c}ar}}]{machielse_quantum_2019}%
  \BibitemOpen
  \bibfield  {author} {\bibinfo {author} {\bibfnamefont {B.}~\bibnamefont {Machielse}}, \bibinfo {author} {\bibfnamefont {S.}~\bibnamefont {Bogdanovic}}, \bibinfo {author} {\bibfnamefont {S.}~\bibnamefont {Meesala}}, \bibinfo {author} {\bibfnamefont {S.}~\bibnamefont {Gauthier}}, \bibinfo {author} {\bibfnamefont {M.}~\bibnamefont {Burek}}, \bibinfo {author} {\bibfnamefont {G.}~\bibnamefont {Joe}}, \bibinfo {author} {\bibfnamefont {M.}~\bibnamefont {Chalupnik}}, \bibinfo {author} {\bibfnamefont {Y.}~\bibnamefont {Sohn}}, \bibinfo {author} {\bibfnamefont {J.}~\bibnamefont {Holzgrafe}}, \bibinfo {author} {\bibfnamefont {R.}~\bibnamefont {Evans}}, \bibinfo {author} {\bibfnamefont {C.}~\bibnamefont {Chia}}, \bibinfo {author} {\bibfnamefont {H.}~\bibnamefont {Atikian}}, \bibinfo {author} {\bibfnamefont {M.}~\bibnamefont {Bhaskar}}, \bibinfo {author} {\bibfnamefont {D.}~\bibnamefont {Sukachev}}, \bibinfo {author} {\bibfnamefont {L.}~\bibnamefont {Shao}}, \bibinfo {author} {\bibfnamefont {S.}~\bibnamefont {Maity}},
  \bibinfo {author} {\bibfnamefont {M.}~\bibnamefont {Lukin}},\ and\ \bibinfo {author} {\bibfnamefont {M.}~\bibnamefont {Lon{\v c}ar}},\ }\bibfield  {title} {\bibinfo {title} {Quantum {Interference} of {Electromechanically} {Stabilized} {Emitters} in {Nanophotonic} {Devices}},\ }\href {https://doi.org/10.1103/PhysRevX.9.031022} {\bibfield  {journal} {\bibinfo  {journal} {Physical Review X}\ }\textbf {\bibinfo {volume} {9}},\ \bibinfo {pages} {031022} (\bibinfo {year} {2019})}\BibitemShut {NoStop}%
\bibitem [{\citenamefont {Wan}\ \emph {et~al.}(2020)\citenamefont {Wan}, \citenamefont {Lu}, \citenamefont {Chen}, \citenamefont {Walsh}, \citenamefont {Trusheim}, \citenamefont {De~Santis}, \citenamefont {Bersin}, \citenamefont {Harris}, \citenamefont {Mouradian}, \citenamefont {Christen}, \citenamefont {Bielejec},\ and\ \citenamefont {Englund}}]{wan_large-scale_2020}%
  \BibitemOpen
  \bibfield  {author} {\bibinfo {author} {\bibfnamefont {N.~H.}\ \bibnamefont {Wan}}, \bibinfo {author} {\bibfnamefont {T.-J.}\ \bibnamefont {Lu}}, \bibinfo {author} {\bibfnamefont {K.~C.}\ \bibnamefont {Chen}}, \bibinfo {author} {\bibfnamefont {M.~P.}\ \bibnamefont {Walsh}}, \bibinfo {author} {\bibfnamefont {M.~E.}\ \bibnamefont {Trusheim}}, \bibinfo {author} {\bibfnamefont {L.}~\bibnamefont {De~Santis}}, \bibinfo {author} {\bibfnamefont {E.~A.}\ \bibnamefont {Bersin}}, \bibinfo {author} {\bibfnamefont {I.~B.}\ \bibnamefont {Harris}}, \bibinfo {author} {\bibfnamefont {S.~L.}\ \bibnamefont {Mouradian}}, \bibinfo {author} {\bibfnamefont {I.~R.}\ \bibnamefont {Christen}}, \bibinfo {author} {\bibfnamefont {E.~S.}\ \bibnamefont {Bielejec}},\ and\ \bibinfo {author} {\bibfnamefont {D.}~\bibnamefont {Englund}},\ }\bibfield  {title} {\bibinfo {title} {Large-scale integration of artificial atoms in hybrid photonic circuits},\ }\href {https://doi.org/10.1038/s41586-020-2441-3} {\bibfield  {journal} {\bibinfo  {journal}
  {Nature}\ }\textbf {\bibinfo {volume} {583}},\ \bibinfo {pages} {226} (\bibinfo {year} {2020})},\ \bibinfo {note} {publisher: Nature Publishing Group}\BibitemShut {NoStop}%
\bibitem [{\citenamefont {Kambs}\ and\ \citenamefont {Becher}(2018)}]{kambs_limitations_2018}%
  \BibitemOpen
  \bibfield  {author} {\bibinfo {author} {\bibfnamefont {B.}~\bibnamefont {Kambs}}\ and\ \bibinfo {author} {\bibfnamefont {C.}~\bibnamefont {Becher}},\ }\bibfield  {title} {\bibinfo {title} {Limitations on the indistinguishability of photons from remote solid state sources},\ }\href {https://doi.org/10.1088/1367-2630/aaea99} {\bibfield  {journal} {\bibinfo  {journal} {New Journal of Physics}\ }\textbf {\bibinfo {volume} {20}},\ \bibinfo {pages} {115003} (\bibinfo {year} {2018})}\BibitemShut {NoStop}%
\bibitem [{\citenamefont {Pittman}\ \emph {et~al.}(1996)\citenamefont {Pittman}, \citenamefont {Strekalov}, \citenamefont {Migdall}, \citenamefont {Rubin}, \citenamefont {Sergienko},\ and\ \citenamefont {Shih}}]{pittman_can_1996}%
  \BibitemOpen
  \bibfield  {author} {\bibinfo {author} {\bibfnamefont {T.~B.}\ \bibnamefont {Pittman}}, \bibinfo {author} {\bibfnamefont {D.~V.}\ \bibnamefont {Strekalov}}, \bibinfo {author} {\bibfnamefont {A.}~\bibnamefont {Migdall}}, \bibinfo {author} {\bibfnamefont {M.~H.}\ \bibnamefont {Rubin}}, \bibinfo {author} {\bibfnamefont {A.~V.}\ \bibnamefont {Sergienko}},\ and\ \bibinfo {author} {\bibfnamefont {Y.~H.}\ \bibnamefont {Shih}},\ }\bibfield  {title} {\bibinfo {title} {Can {Two}-{Photon} {Interference} be {Considered} the {Interference} of {Two} {Photons}?},\ }\href {https://doi.org/10.1103/PhysRevLett.77.1917} {\bibfield  {journal} {\bibinfo  {journal} {Physical Review Letters}\ }\textbf {\bibinfo {volume} {77}},\ \bibinfo {pages} {1917} (\bibinfo {year} {1996})}\BibitemShut {NoStop}%
\bibitem [{\citenamefont {Legero}\ \emph {et~al.}(2003)\citenamefont {Legero}, \citenamefont {Wilk}, \citenamefont {Kuhn},\ and\ \citenamefont {Rempe}}]{legero_time-resolved_2003}%
  \BibitemOpen
  \bibfield  {author} {\bibinfo {author} {\bibfnamefont {T.}~\bibnamefont {Legero}}, \bibinfo {author} {\bibfnamefont {T.}~\bibnamefont {Wilk}}, \bibinfo {author} {\bibfnamefont {A.}~\bibnamefont {Kuhn}},\ and\ \bibinfo {author} {\bibfnamefont {G.}~\bibnamefont {Rempe}},\ }\bibfield  {title} {\bibinfo {title} {Time-resolved two-photon quantum interference},\ }\href {https://doi.org/10.1007/s00340-003-1337-x} {\bibfield  {journal} {\bibinfo  {journal} {Applied Physics B}\ }\textbf {\bibinfo {volume} {77}},\ \bibinfo {pages} {797} (\bibinfo {year} {2003})}\BibitemShut {NoStop}%
\bibitem [{\citenamefont {Halder}\ \emph {et~al.}(2007)\citenamefont {Halder}, \citenamefont {Beveratos}, \citenamefont {Gisin}, \citenamefont {Scarani}, \citenamefont {Simon},\ and\ \citenamefont {Zbinden}}]{halder_entangling_2007}%
  \BibitemOpen
  \bibfield  {author} {\bibinfo {author} {\bibfnamefont {M.}~\bibnamefont {Halder}}, \bibinfo {author} {\bibfnamefont {A.}~\bibnamefont {Beveratos}}, \bibinfo {author} {\bibfnamefont {N.}~\bibnamefont {Gisin}}, \bibinfo {author} {\bibfnamefont {V.}~\bibnamefont {Scarani}}, \bibinfo {author} {\bibfnamefont {C.}~\bibnamefont {Simon}},\ and\ \bibinfo {author} {\bibfnamefont {H.}~\bibnamefont {Zbinden}},\ }\bibfield  {title} {\bibinfo {title} {Entangling independent photons by time measurement},\ }\href {https://doi.org/10.1038/nphys700} {\bibfield  {journal} {\bibinfo  {journal} {Nature Physics}\ }\textbf {\bibinfo {volume} {3}},\ \bibinfo {pages} {692} (\bibinfo {year} {2007})},\ \bibinfo {note} {publisher: Nature Publishing Group}\BibitemShut {NoStop}%
\bibitem [{\citenamefont {Wheeler}(1978)}]{wheeler_past_1978}%
  \BibitemOpen
  \bibfield  {author} {\bibinfo {author} {\bibfnamefont {J.~A.}\ \bibnamefont {Wheeler}},\ }\bibfield  {title} {\bibinfo {title} {The ``{Past}'' and the ``{Delayed}-{Choice}'' {Double}-{Slit} {Experiment}},\ }in\ \href {https://doi.org/10.1016/B978-0-12-473250-6.50006-6} {\emph {\bibinfo {booktitle} {Mathematical {Foundations} of {Quantum} {Theory}}}},\ \bibinfo {editor} {edited by\ \bibinfo {editor} {\bibfnamefont {A.~R.}\ \bibnamefont {Marlow}}}\ (\bibinfo  {publisher} {Academic Press},\ \bibinfo {year} {1978})\ pp.\ \bibinfo {pages} {9--48}\BibitemShut {NoStop}%
\bibitem [{\citenamefont {Zhao}\ \emph {et~al.}(2014)\citenamefont {Zhao}, \citenamefont {Zhang}, \citenamefont {Yang}, \citenamefont {Sang}, \citenamefont {Jiang}, \citenamefont {Bao},\ and\ \citenamefont {Pan}}]{zhao_entangling_2014}%
  \BibitemOpen
  \bibfield  {author} {\bibinfo {author} {\bibfnamefont {T.-M.}\ \bibnamefont {Zhao}}, \bibinfo {author} {\bibfnamefont {H.}~\bibnamefont {Zhang}}, \bibinfo {author} {\bibfnamefont {J.}~\bibnamefont {Yang}}, \bibinfo {author} {\bibfnamefont {Z.-R.}\ \bibnamefont {Sang}}, \bibinfo {author} {\bibfnamefont {X.}~\bibnamefont {Jiang}}, \bibinfo {author} {\bibfnamefont {X.-H.}\ \bibnamefont {Bao}},\ and\ \bibinfo {author} {\bibfnamefont {J.-W.}\ \bibnamefont {Pan}},\ }\bibfield  {title} {\bibinfo {title} {Entangling {Different}-{Color} {Photons} via {Time}-{Resolved} {Measurement} and {Active} {Feed} {Forward}},\ }\href {https://doi.org/10.1103/PhysRevLett.112.103602} {\bibfield  {journal} {\bibinfo  {journal} {Physical Review Letters}\ }\textbf {\bibinfo {volume} {112}},\ \bibinfo {pages} {103602} (\bibinfo {year} {2014})}\BibitemShut {NoStop}%
\bibitem [{\citenamefont {Parker}\ \emph {et~al.}(2024)\citenamefont {Parker}, \citenamefont {Arjona~Mart{\'\i}nez}, \citenamefont {Chen}, \citenamefont {Stramma}, \citenamefont {Harris}, \citenamefont {Michaels}, \citenamefont {Trusheim}, \citenamefont {Hayhurst~Appel}, \citenamefont {Purser}, \citenamefont {Roth}, \citenamefont {Englund},\ and\ \citenamefont {Atat{\"u}re}}]{parker_diamond_2024}%
  \BibitemOpen
  \bibfield  {author} {\bibinfo {author} {\bibfnamefont {R.~A.}\ \bibnamefont {Parker}}, \bibinfo {author} {\bibfnamefont {J.}~\bibnamefont {Arjona~Mart{\'\i}nez}}, \bibinfo {author} {\bibfnamefont {K.~C.}\ \bibnamefont {Chen}}, \bibinfo {author} {\bibfnamefont {A.~M.}\ \bibnamefont {Stramma}}, \bibinfo {author} {\bibfnamefont {I.~B.}\ \bibnamefont {Harris}}, \bibinfo {author} {\bibfnamefont {C.~P.}\ \bibnamefont {Michaels}}, \bibinfo {author} {\bibfnamefont {M.~E.}\ \bibnamefont {Trusheim}}, \bibinfo {author} {\bibfnamefont {M.}~\bibnamefont {Hayhurst~Appel}}, \bibinfo {author} {\bibfnamefont {C.~M.}\ \bibnamefont {Purser}}, \bibinfo {author} {\bibfnamefont {W.~G.}\ \bibnamefont {Roth}}, \bibinfo {author} {\bibfnamefont {D.}~\bibnamefont {Englund}},\ and\ \bibinfo {author} {\bibfnamefont {M.}~\bibnamefont {Atat{\"u}re}},\ }\bibfield  {title} {\bibinfo {title} {A diamond nanophotonic interface with an optically accessible deterministic electronuclear spin register},\ }\href
  {https://doi.org/10.1038/s41566-023-01332-8} {\bibfield  {journal} {\bibinfo  {journal} {Nature Photonics}\ }\textbf {\bibinfo {volume} {18}},\ \bibinfo {pages} {156} (\bibinfo {year} {2024})},\ \bibinfo {note} {publisher: Nature Publishing Group}\BibitemShut {NoStop}%
\bibitem [{\citenamefont {Beugnon}\ \emph {et~al.}(2006)\citenamefont {Beugnon}, \citenamefont {Jones}, \citenamefont {Dingjan}, \citenamefont {Darqui{\'e}}, \citenamefont {Messin}, \citenamefont {Browaeys},\ and\ \citenamefont {Grangier}}]{beugnon_quantum_2006}%
  \BibitemOpen
  \bibfield  {author} {\bibinfo {author} {\bibfnamefont {J.}~\bibnamefont {Beugnon}}, \bibinfo {author} {\bibfnamefont {M.~P.~A.}\ \bibnamefont {Jones}}, \bibinfo {author} {\bibfnamefont {J.}~\bibnamefont {Dingjan}}, \bibinfo {author} {\bibfnamefont {B.}~\bibnamefont {Darqui{\'e}}}, \bibinfo {author} {\bibfnamefont {G.}~\bibnamefont {Messin}}, \bibinfo {author} {\bibfnamefont {A.}~\bibnamefont {Browaeys}},\ and\ \bibinfo {author} {\bibfnamefont {P.}~\bibnamefont {Grangier}},\ }\bibfield  {title} {\bibinfo {title} {Quantum interference between two single photons emitted by independently trapped atoms},\ }\href {https://doi.org/10.1038/nature04628} {\bibfield  {journal} {\bibinfo  {journal} {Nature}\ }\textbf {\bibinfo {volume} {440}},\ \bibinfo {pages} {779} (\bibinfo {year} {2006})}\BibitemShut {NoStop}%
\bibitem [{\citenamefont {Binder}\ \emph {et~al.}(2017)\citenamefont {Binder}, \citenamefont {Stark}, \citenamefont {Tomek}, \citenamefont {Scheuer}, \citenamefont {Frank}, \citenamefont {Jahnke}, \citenamefont {M{\"u}ller}, \citenamefont {Schmitt}, \citenamefont {Metsch}, \citenamefont {Unden}, \citenamefont {Gehring}, \citenamefont {Huck}, \citenamefont {Andersen}, \citenamefont {Rogers},\ and\ \citenamefont {Jelezko}}]{binder_qudi_2017}%
  \BibitemOpen
  \bibfield  {author} {\bibinfo {author} {\bibfnamefont {J.~M.}\ \bibnamefont {Binder}}, \bibinfo {author} {\bibfnamefont {A.}~\bibnamefont {Stark}}, \bibinfo {author} {\bibfnamefont {N.}~\bibnamefont {Tomek}}, \bibinfo {author} {\bibfnamefont {J.}~\bibnamefont {Scheuer}}, \bibinfo {author} {\bibfnamefont {F.}~\bibnamefont {Frank}}, \bibinfo {author} {\bibfnamefont {K.~D.}\ \bibnamefont {Jahnke}}, \bibinfo {author} {\bibfnamefont {C.}~\bibnamefont {M{\"u}ller}}, \bibinfo {author} {\bibfnamefont {S.}~\bibnamefont {Schmitt}}, \bibinfo {author} {\bibfnamefont {M.~H.}\ \bibnamefont {Metsch}}, \bibinfo {author} {\bibfnamefont {T.}~\bibnamefont {Unden}}, \bibinfo {author} {\bibfnamefont {T.}~\bibnamefont {Gehring}}, \bibinfo {author} {\bibfnamefont {A.}~\bibnamefont {Huck}}, \bibinfo {author} {\bibfnamefont {U.~L.}\ \bibnamefont {Andersen}}, \bibinfo {author} {\bibfnamefont {L.~J.}\ \bibnamefont {Rogers}},\ and\ \bibinfo {author} {\bibfnamefont {F.}~\bibnamefont {Jelezko}},\ }\bibfield  {title} {\bibinfo {title}
  {Qudi: {A} modular python suite for experiment control and data processing},\ }\href {https://doi.org/10.1016/j.softx.2017.02.001} {\bibfield  {journal} {\bibinfo  {journal} {SoftwareX}\ }\textbf {\bibinfo {volume} {6}},\ \bibinfo {pages} {85} (\bibinfo {year} {2017})}\BibitemShut {NoStop}%
\end{thebibliography}%

\end{document}